\documentclass{PoS}

\usepackage{multirow}
\usepackage{sidecap}
\usepackage{adjustbox}
\usepackage{fixltx2e}
\MakeRobust{\ref}

\def\si{^1 \hskip -0.025in S _0}

\newcommand{\gsim}{\raisebox{-0.7ex}{$\stackrel{\textstyle >}{\sim}$ }}

\title{Nuclear Physics}

\ShortTitle{Nuclear Physics}

\author{\speaker{Martin J. Savage}%
         \thanks{I would like to thank all members of the NPLQCD collaboration.}\\
        Institute For Nuclear Theory, University of Washington, Seattle, WA 98195-1550\\
        E-mail: \email{mjs5@uw.edu}}

\abstract{Lattice QCD is making good progress toward calculating the structure and 
properties of light nuclei and the forces between nucleons. These calculations will ultimately 
refine the nuclear forces, particularly in the three- and four-nucleon sector and the 
short-distance interactions of nucleons with electroweak currents, and allow for a reduction of 
uncertainties in nuclear many-body calculations of nuclei and their reactions.  
After highlighting their importance, particularly to the Nuclear Physics and High-Energy Physics 
experimental programs, I will discuss the progress that has been made toward 
achieving these goals and the challenges that remain.
}

\FullConference{34th annual International Symposium on Lattice Field Theory\\
24-30 July 2016\\
University of Southampton, UK}

\begin{document}

\section{Introduction}

Nuclear physics is on the brink of being changed in remarkable ways by the use of Lattice QCD (LQCD) to provide 
reliable calculations of low-energy strong interaction processes that cannot be reliably obtained by any other means.
The crucial step of verifying  LQCD as a  rigorous source of strong-interaction observables, complementary to experiment,  
is beginning to happen by the reproduction of nuclear physics quantities that are already  precisely known.
As this verification stage  is in process, genuine predictions of QCD performed with LQCD are beginning to emerge.
There are many important strong interaction quantities, impacting a broad array of research programs and technologies, 
that are required to be more precisely known than will be possible experimentally or through  known analytical theory methods.
One class of such quantities is the multi-nucleon forces, such as the three-neutron and four-neutron interactions.
\begin{figure}
\begin{center}
\includegraphics[width=0.4\linewidth]{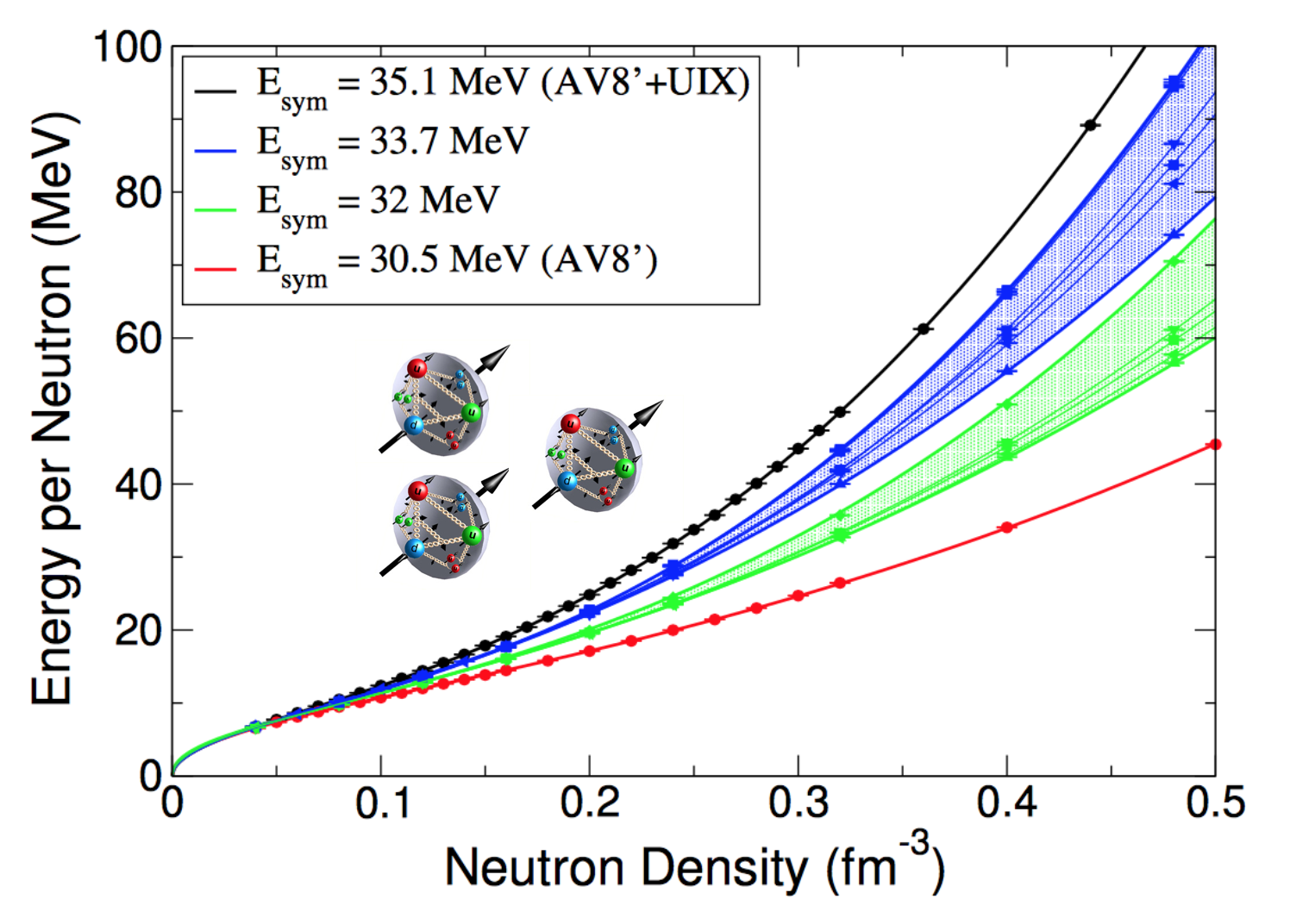}
\ \ \ \ \ \ \ \ \ \ 
\raisebox{0.05\height}{ \includegraphics[width=0.3\linewidth]{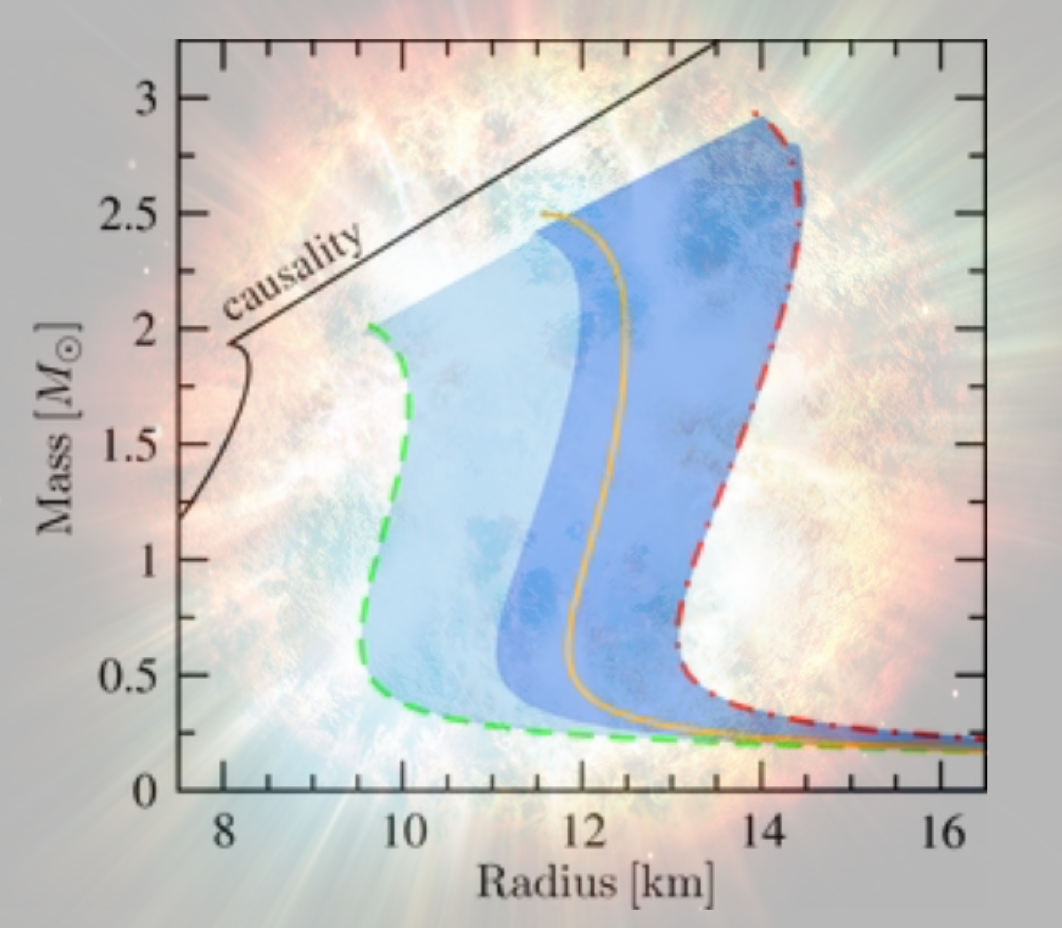} }
\end{center}
\caption{
The energy per neutron as a function of neutron density for a range of  parameters defining nuclear forces 
(left panel)~\protect\cite{Gandolfi:2015jma}, 
and the 
allowed region of mass and radius of neutron stars for a similar range parameters (right panel)~\protect\cite{Hebeler:2013nza}.
\label{fig:neutronmatter}}
\end{figure}
Figure~\ref{fig:neutronmatter} shows the present uncertainties in the energy per neutron as a function of neutron density 
from current nuclear forces, which is seen to become significant beyond nuclear matter densities~\cite{Gandolfi:2015jma}.  
This, and other uncertainties, impact our ability to calculate basic properties of neutron stars, such as the 
mass-radius relation~\cite{Hebeler:2013nza},  also shown in Figure~\ref{fig:neutronmatter}.
The US is building FRIB (Facility for Rare Isotope Beams), and other countries are building similar facilities, to 
study nuclei and nuclear systems that exist for a very short time.  
These nuclei participate in explosive astrophysical environments, and have so far proven difficult to examine in the laboratory.
FRIB, currently under construction at Michigan State University, 
will make major inroads into the detailed study of these exotic systems.
The anticipated measurements will  
refine our understanding of multi-neutron forces, particularly the three-neutron interaction, and more generally three-body and four-body forces.
LQCD calculations of few-nucleon systems will, combined with nuclear many-body calculations, will provide complementary information and constraints on the multi-nucleon forces.
The discovery of gravitational waves emitted from inspiraling black-hole binary systems~\cite{Abbott:2016blz}
 was a major accomplishment, opening a new era of exploration of the universe,
 and the gravitational wave signals expected from inspiring binary neutron star systems 
 will be sensitive to the nuclear equation of state (EoS), and hence the three-neutron and higher forces.

A second class of quantities that need to be known with higher precision  are 
 the interactions of nuclear systems with electroweak probes, and candidate particles beyond the standard model  (BSM).  
 Such processes are critical to much of the US's nuclear and particle physics experimental programs,
 including the planned ton-scale $0\nu\beta\beta$-decay experiment, the Deep Underground Neutrino Experiment (DUNE), neutron electric dipole moment searches, and cross sections for simple fusion processes.
\begin{figure}
\begin{center}
\includegraphics[width=0.65\linewidth]{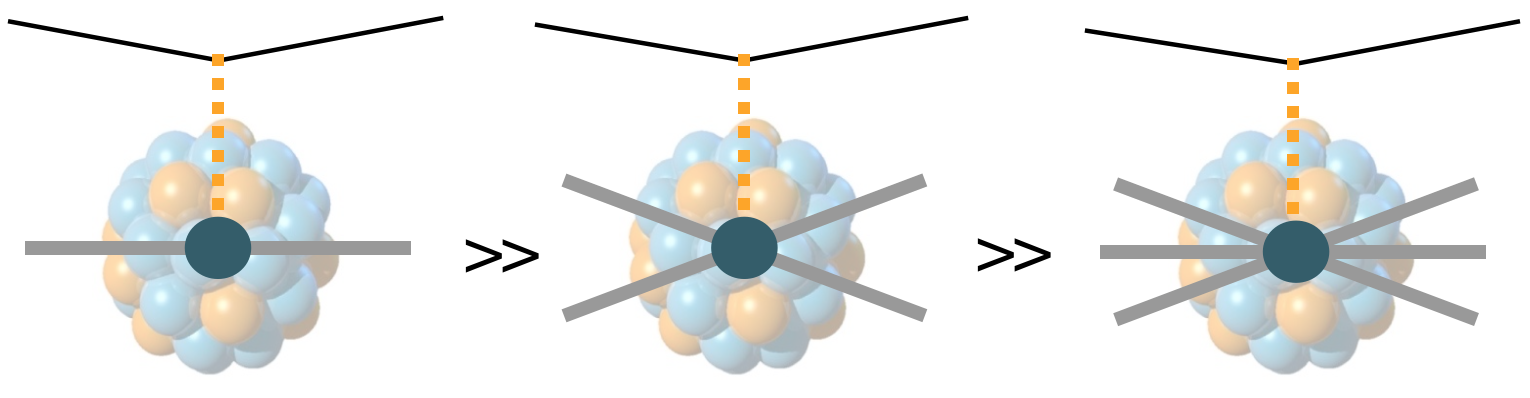}
\end{center}
\caption{
A cartoon of the relative importance of multi-nucleon interactions with external probes.
\label{fig:multicurrents}
}
\end{figure}
For electromagnetic interactions, there is a well-established hierarchy for the relative contributions of multi-nucleon effects, as shown in Figure~\ref{fig:multicurrents}.
Low-energy processes are dominated by one-body interactions, with two-body interactions typically at the few percent level, and three-nucleon interactions further suppressed.
For nuclear matrix elements of the axial current,
calculations with truncated model spaces require a reduced (quenched) value of $g_A$ for the one-body interaction.
This is most likely due to inconsistent treatments in matching from chiral effective field theories to the nuclear many-body space, where multi-nucleon interactions with the axial current will likely compensate the quenched $g_A$.
This highlights the importance of appreciating that a nucleus is not simply a collection of non-interacting nucleons, but a complex system with multi-nucleon interactions, including with external probes.

\begin{figure}
\begin{center}
\includegraphics[width=0.35\linewidth]{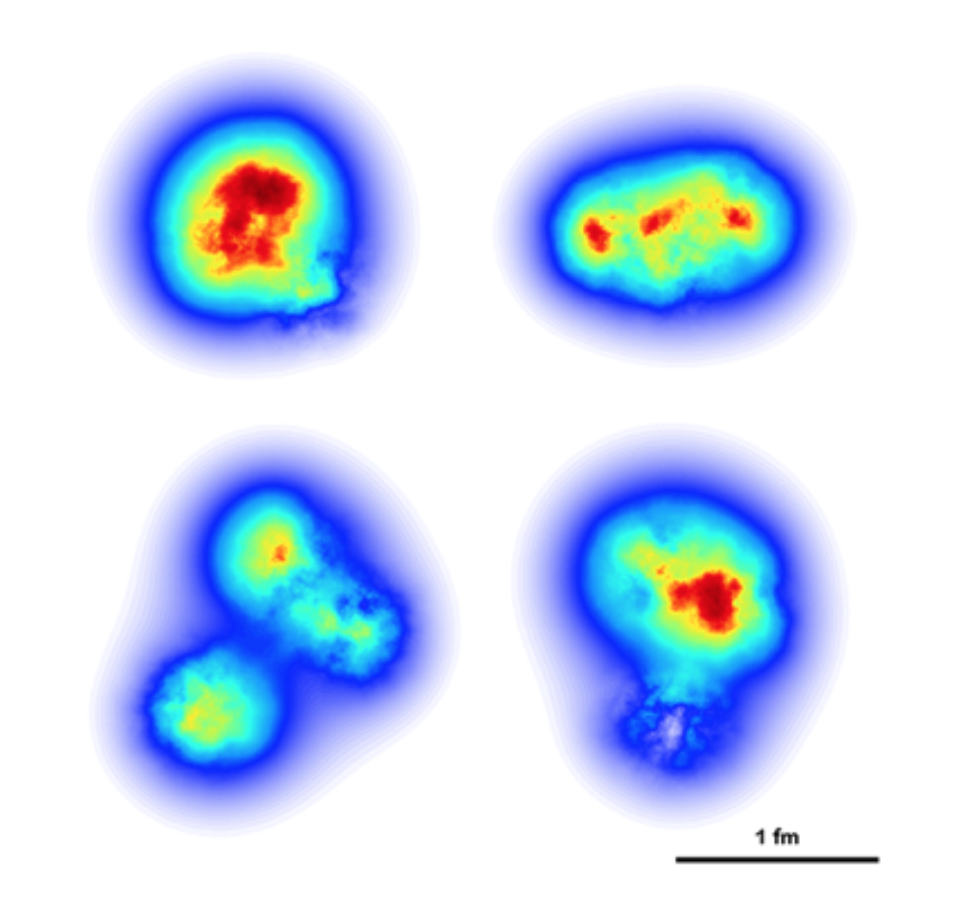}
\end{center}
\caption{
Time-dependence of the gluon field in a nucleus~\protect\cite{Mantysaari:2016ykx}. 
\label{fig:gluons}}
\end{figure}
A third class of quantities is directly related to the US's planned electron-ion collider (EIC) that is 
designed to precisely measure the gluonic structure of the nucleon and of nuclei.
This was identified as a long-term priority for nuclear physics~\cite{nsacLRP2015},
which may lead to an EIC  in the United States.
LQCD gluonic calculations are notoriously difficult, but are essential to the success of an EIC program.
Figure~\ref{fig:gluons} shows a classical calculation of the time dependence of the gluon density in a 
nucleus~\cite{Mantysaari:2016ykx}.

\begin{figure}
\begin{center}
\includegraphics[width=0.65\linewidth]{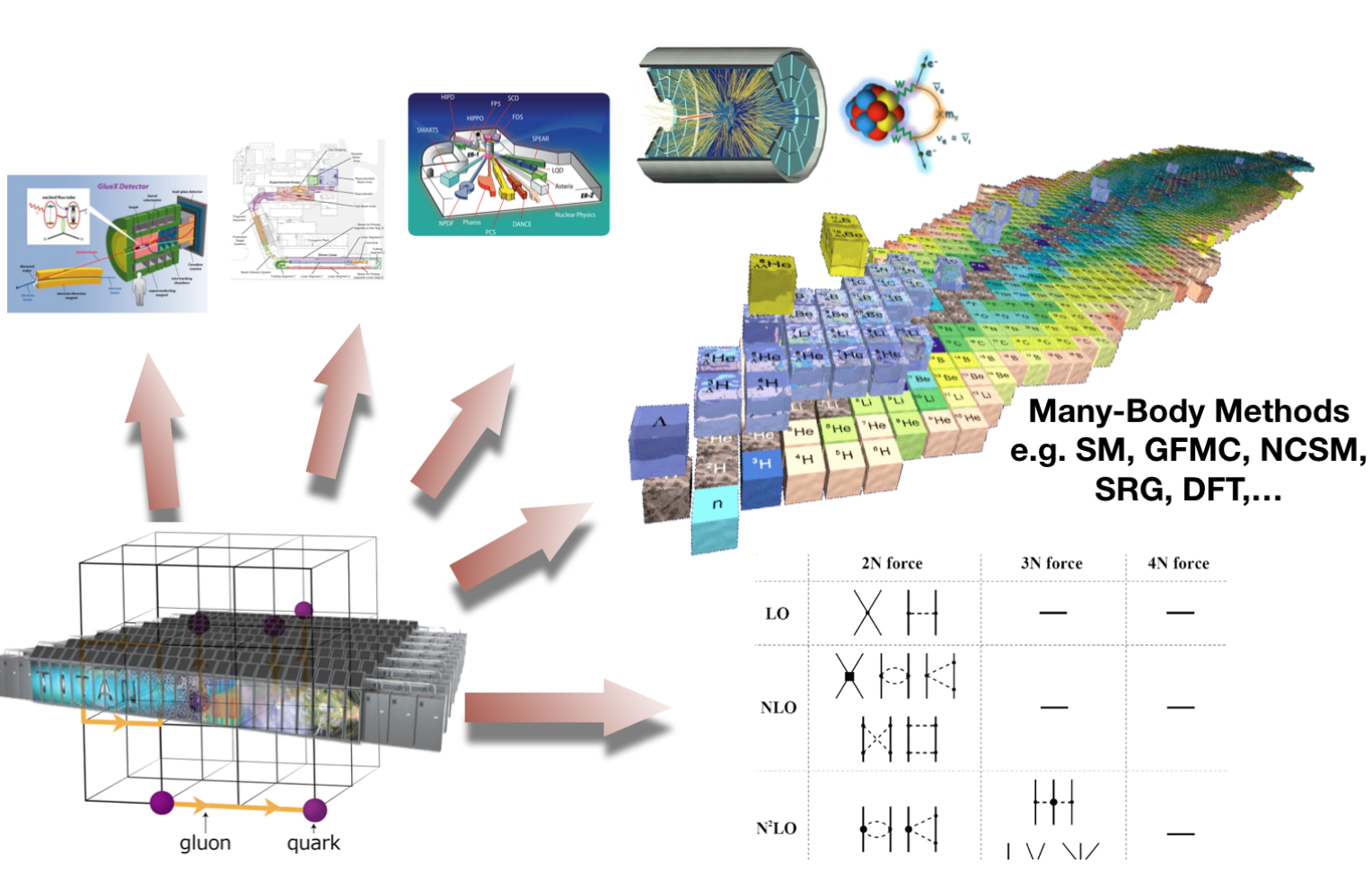}
\end{center}
\caption{
Cartoon of the grand plan for LQCD in nuclear physics. 
}
\label{fig:GrandPlan}
\end{figure}

The grand plan for using the numerical technique of LQCD in nuclear physics is quite simple.
 It is to be expected that systems involving up to twelve nucleons or 
so will be accessible to LQCD calculations in the not so distant future. 
For small and modest lattice volumes, the energy splittings between levels is 
sufficient so that there is the 
possibility of isolating them with techniques such as the variational method,
and  
the binding energies, and  more generally observables associated with the ground states and low-lying excited states, can be  compared with experiment.  
More generally, the energy eigenvalues computed in a range of lattice volumes
will be used to refine  effective nuclear (many-body) forces through appropriate 
finite-volume matching calculations
that can then be used to make predictions in Minkowski space, as sketched in Figure~\ref{fig:GrandPlan}.

One of the challenges facing nucleon and multi-nucleon systems is the signal-to-noise problem.
At asymptotically large times, the even moments of a nucleus correlation function is dominated by 
multi-pion states, which fall exponentially with  a multiple of the  pion mass, while the odd moments 
are suppressed by at least one factor of the nucleon mass.
At short times, for appropriate source structures, all moments of the correlation functions are determined 
by the nucleon mass.
At intermediate times,  the signal-to-noise ratio is degrading but with an exponent that is significantly less than the canonical 
$A( M_N-{3\over 2} m_\pi)$~\cite{Beane:2009kya}, and it is in such an intermediate
time interval (``Golden Window'') that plateaus in multi-nucleon correlation functions can be identified, 
and binding energies and scattering parameters determined.

Until about five years ago, the quark contractions required to form nuclear correlation functions required 
excessive computational resources to evaluate.
For light nuclear systems, the contraction problem was solved through understanding the symmetry of the contractions, implementing recursion algorithms, and automated code generation.
This reduced the impact of the quark contractions on the overall computational resources required for production of s-shell nuclei and hypernuclei.~\cite{Detmold:2012eu,Doi:2012xd,Yamazaki:2012hi}.

The methodology for extracting phase-shifts, scattering parameters and two-nucleon bound states from LQCD calculations 
was formulated in detail by Luscher~\cite{Luscher:1986pf,Luscher:1990ux}.
The two-hadron wave function, $\psi(r)$ satisfies 
\begin{eqnarray}
-{1\over 2\mu}\Delta \psi(r) 
\ +\ 
{1\over 2} \int d^3{\bf r}^\prime\ U_E({\bf r},{\bf r}^\prime) \psi(r^\prime) 
& = & E  \psi(r) 
\ \ \ ,
\label{eq:betasalpeter}
\end{eqnarray}
where $U_E({\bf r},{\bf r}^\prime)$ is an energy-dependent non-local ``potential'', and where the total energy of the system is 
$W=2\sqrt{M^2 + M E}$. 
$U_E({\bf r},{\bf r}^\prime)$ depends analytically on $E$ in the region below the inelastic threshold 
and is a smooth function of the coordinates.
This construction readily reveals the L\"uscher relations which, after truncation in angular momentum space, 
relates  energy splittings (to non-interacting states) 
to the parameters defining the 
S-matrix in that system.

The HALQCD collaboration uses wall sources to create two-nucleon states and forms correlation functions that depend upon spatial separation by a composite product sink comprised 
of two nucleon field operators, $G_{NN}(r)=Z_{NN}(r) e^{i {\bf p}\cdot{\bf R}} \psi(r)$.
At the energy eigenvalues of the lattice calculation, this object satisfies the Schrodinger equation and a resulting potential,
$U_E(r)$, 
can be  identified~\cite{HALQCD:2012aa}.
HALQCD has been performing calculations at the physical point and have derived such
$U_E(r)$'s.  
The interpretation of the derived $U_E(r)$'s remains to be determined as they do not have sufficient statistics to identify plateaus in their correlation functions.
\begin{figure}
\begin{center}
\includegraphics[width=0.45\linewidth]{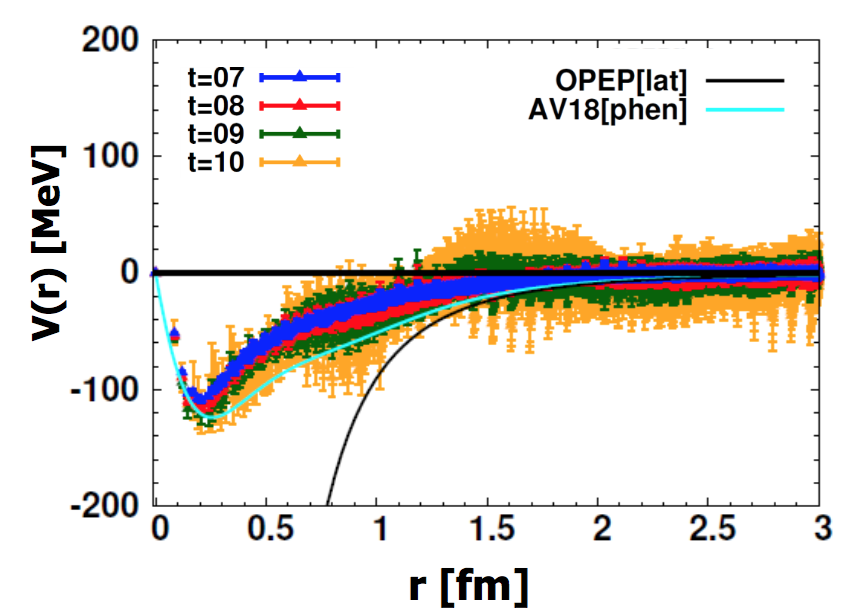}
\end{center}
\caption{
A  tensor $U_E(r)$ derived from correlation functions in the $^3S_1$-$^3D_1$ coupled channels.
[I thank Takumi Doi for providing this figure.]
\label{fig:tensorHALQCD}}
\end{figure}
The single nucleon correlation function from these wall sources plateaus after 
$t\sim 18$, and  the $U_E(r)$'s  in the two-nucleon sector, 
an example of which is shown in Figure~\ref{fig:tensorHALQCD},
are  contaminated by excited states.
It would be helpful for the HALQCD collaboration to extract the masses associated 
with the long-distance behavior of the $U_E(r)$'s.

A well-defined procedure for extracting scattering parameters and S-matrix elements from  
two-hadron energy eigenvalues is  L\"usher's method,
which has been used extensively to study hadron-hadron interactions.
The PACS, NPLQCD and Mainz collaborations use  effective masses to directly measure the energy eigenvalues of multi-nucleon states in a 
range of lattice volumes, to directly measure binding energies and  use L\"usher's method to extract scattering 
parameters and phase shifts. 
\begin{figure}
\begin{center}
\vskip -0.5in
\includegraphics[angle=270,width=0.5\linewidth]{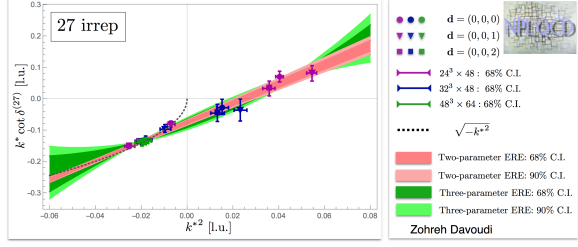}
\end{center}
\vskip -0.7in
\caption{
$k^*\cot\delta$ as a function of  $k^{*2}$ in the {\bf 27} irrep. of flavor SU(3)
calculated at a pion mass of $m_\pi\sim 805~{\rm MeV}$~\protect\cite{Beane:2013br}.
{\bf d} denotes the boost vector in lattice units.  
Two- and three-parameter ERE fits are shown by the shaded regions.
\label{fig:kcot27nplqcd}
}
\end{figure}
In simple systems, by extracting $k^*\cot\delta$ below the inelastic threshold  from systems 
with a range of boosts and 
boundary conditions~\cite{Davoudi:2011md,Briceno:2013lba,Briceno:2013hya}, 
single-channel phase shifts can be determined  from energy eigenvalues
along with scattering parameters in the 
effective range expansion (ERE) which is  applicable below the t-channel cut, 
an example of which is shown  in Figure~\ref{fig:kcot27nplqcd}.
Multi-channel systems can also be addressed, and the corresponding S-matrix elements can be constrained 
from finite-volume  energy eigenvalues,  but the analysis is more complex~\cite{Briceno:2013bda}.


The PACS collaboration has been calculating the binding energies of  s-shell nuclei for a number of years, 
and, impressively,  are now performing calculations at the physical pion mass~\cite{Yamazaki:2012hi,Yamazaki:2015asa}.
\begin{figure}
\begin{center}
\includegraphics[width=0.65\linewidth]{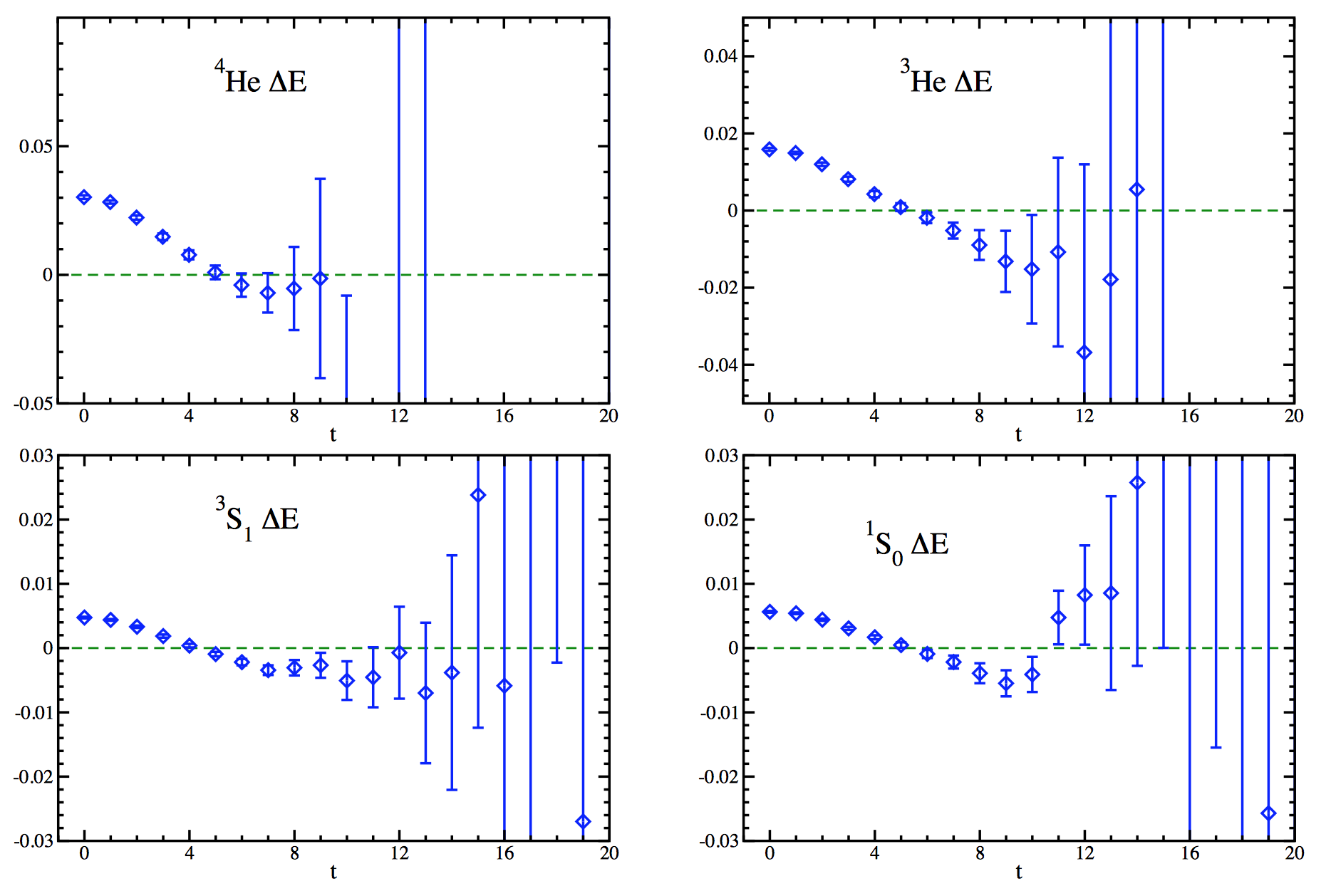}
\end{center}
\caption{
Effective mass plots of the s-shell nuclei calculated at the physical pion mass, 
$m_\pi\sim 140~{\rm MeV}$,
by the PACS collaboration.
[I thank Takeshi Yamasaki for providing these figures.]
\label{fig:PACS140MeV}
}
\end{figure}
Figure~\ref{fig:PACS140MeV} shows the current status of these calculations.   
The two-nucleon systems are showing encouraging signs of developing plateaus.  
Given the lattice volumes currently employed, the finite-volume effects for these levels are expected to be significant~\cite{Beane:2003da} in connecting to experiment.  
It is also  encouraging to see correlation functions for $^3$He and $^4$He, and one hopes that  
increased statistics will reveal the binding energies of these systems that are measured in experiment (modulo electromagnetic and isospin breaking effects).
Unfortunately, it appears that the correlation functions are entering the exponentially-degrading signal-to-noise region at the times when plateaus are forming in $^3$He and $^4$He.  
\begin{figure}
\begin{center}
 \raisebox{0.1\height}{   \includegraphics[width=0.355\linewidth]{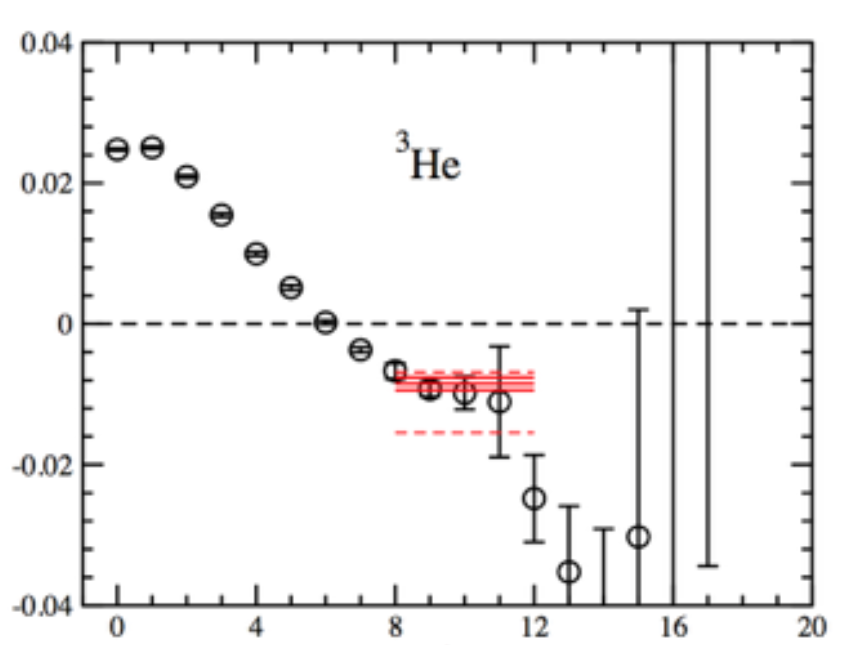}  }\ \ \ \ \ 
\includegraphics[width=0.45\linewidth]{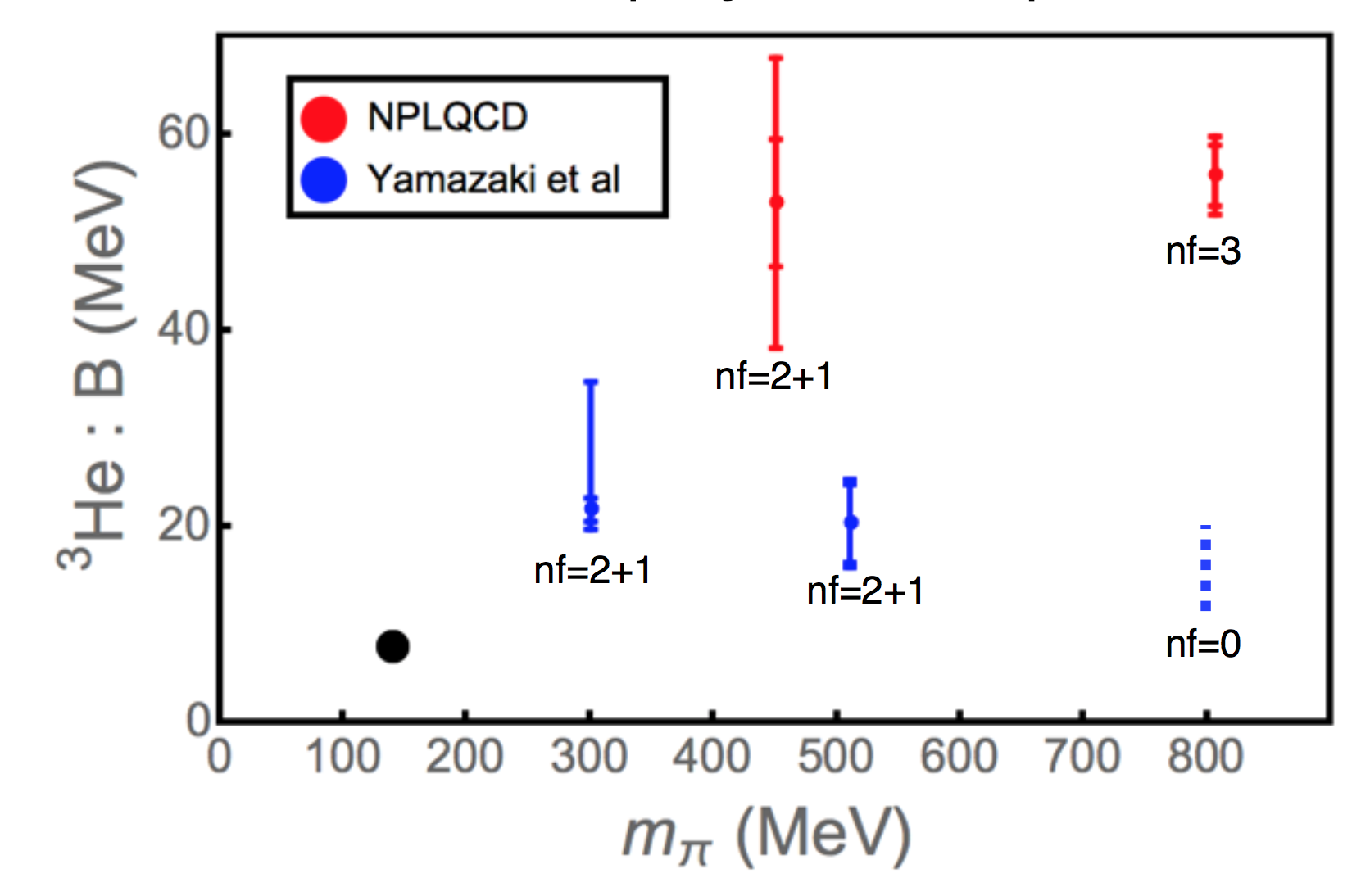}
\end{center}
\caption{
The left panel shows a $^3$He effective mass plot from the PACS collaboration at $m_\pi\sim 300~{\rm MeV}$,
while the right panel is a summary of calculations of $^3$He binding energy as a function of the pion 
mass~\cite{Beane:2012vq,Yamazaki:2015asa,Orginos:2015aya}.
\label{fig:3He}
}
\end{figure}
The PACS correlation functions 
shown in Figure~\ref{fig:3He} are created from smeared sources for the quark propagators, from which the single nucleon correlation functions are observed to plateau around time-slice $t=7$.

\begin{figure}
\begin{center}
 \raisebox{0.1\height}{   \includegraphics[width=0.55\linewidth]{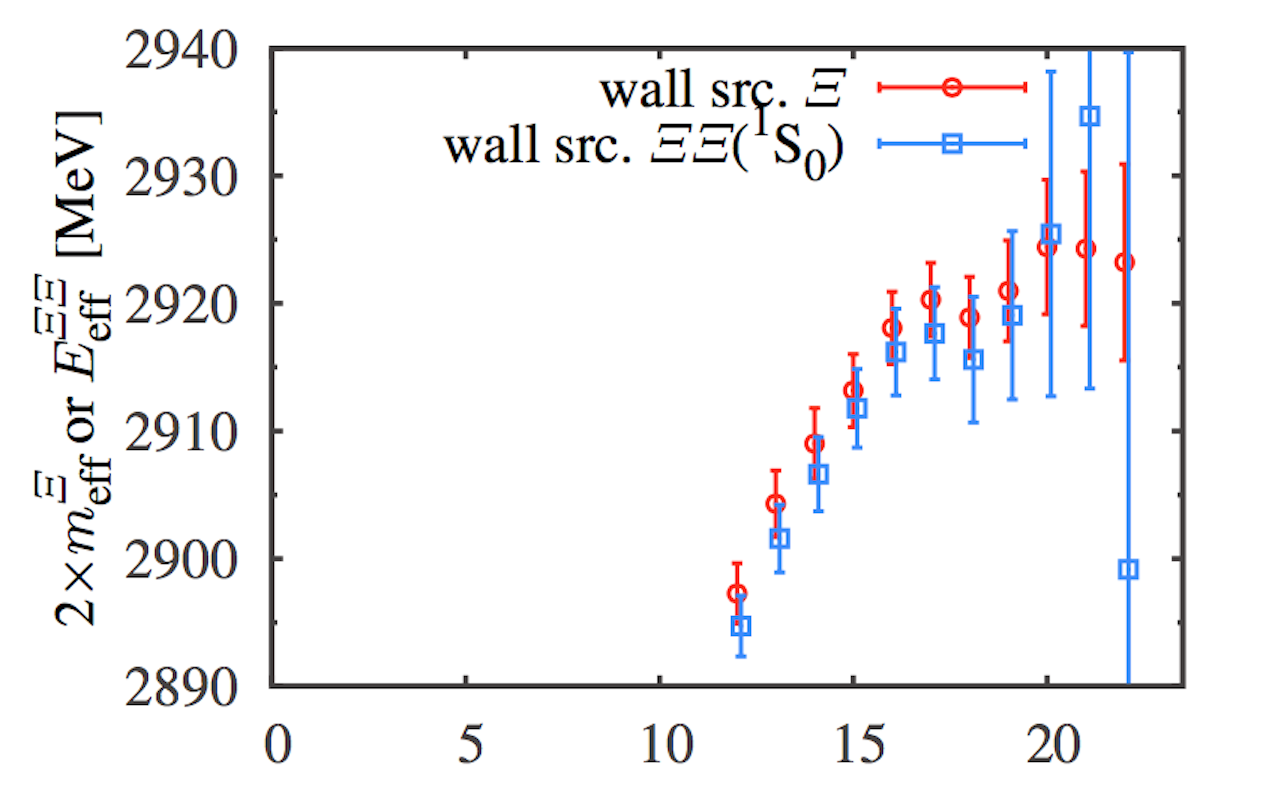}  }\end{center}
\vskip -0.3in
\caption{
Effective mass plots for the $\Xi$ and $\Xi\Xi (\si)$ from wall sources generate by 
HALQCD~\protect\cite{Iritani:2016jie}.
[I thank Takumi Doi for providing this figure.]
\label{fig:HALQCDwall}
}
\end{figure}

HALQCD  has been calculating $U_E(r)$~\cite{HALQCD:2012aa}, Eq.~(\ref{eq:betasalpeter}), 
over a range of pion masses for a large number of two-baryon systems, including NN and hypernuclear systems, for example Refs.~\cite{Sasaki:2016gpc,Ishii:2016zsf}, 
and a small number of three-baryon systems~\cite{Doi:2011gq}.  
The signal-to-noise issues are less 
severe for the heavier systems and consequently the signals for the systems with large strangeness are better 
than for the NN systems.
They have  high precision determinations of the $U_E(r)$ in the H-dibaryon coupled-channels system, 
from which the draw the preliminary conclusion that the H-dibaryon is a resonance above $\Lambda\Lambda$ threshold 
at the physical point.
In deriving these $U_E(r)$, the energy of the two-baryon system is required, and 
HALQCD 
acts with time-derivatives at time slices where the effective mass has not plateau'd and attributes the non-plateau'ing to 
two-baryon excitations in the lattice volume, ignoring the possibility of single baryon excitations.
As the single nucleon effective masses from the wall sources do not plateau until time-slice  
$t\sim 18$, 
it seems that such extractions are contaminated by single nucleon excitations until $t\sim 18$, 
as shown in Figure~\ref{fig:HALQCDwall}~\cite{Iritani:2016jie}, 
and deriving 
$U_E(r)$ in this way introduces  uncertainties for $t < 18$ that are not accounted for.  
The $U_E(r)$ that have been calculated by HALQCD are derived from the $t \leq 10$ range of the correlation functions.
It is very encouraging that HALQCD is performing calculations at the physical point, 
and early results  are shown in Figure~\ref{fig:tensorHALQCD},
and we look forward to them being able to extract 
$U_E(r)$ with precision in time intervals where the single-baryon and two-baryon correlation functions have both 
established plateaus in the effective mass.


The Mainz Lattice collaboration has been calculating the binding energy of two-baryon systems, 
 in the SU(3) limit, 
and including SU(3) breaking, over a range of pion masses~\cite{Green:2014dea,Junnarkar:2015jyf}.   
\begin{figure}
\begin{center}
 \includegraphics[width=0.45\linewidth]{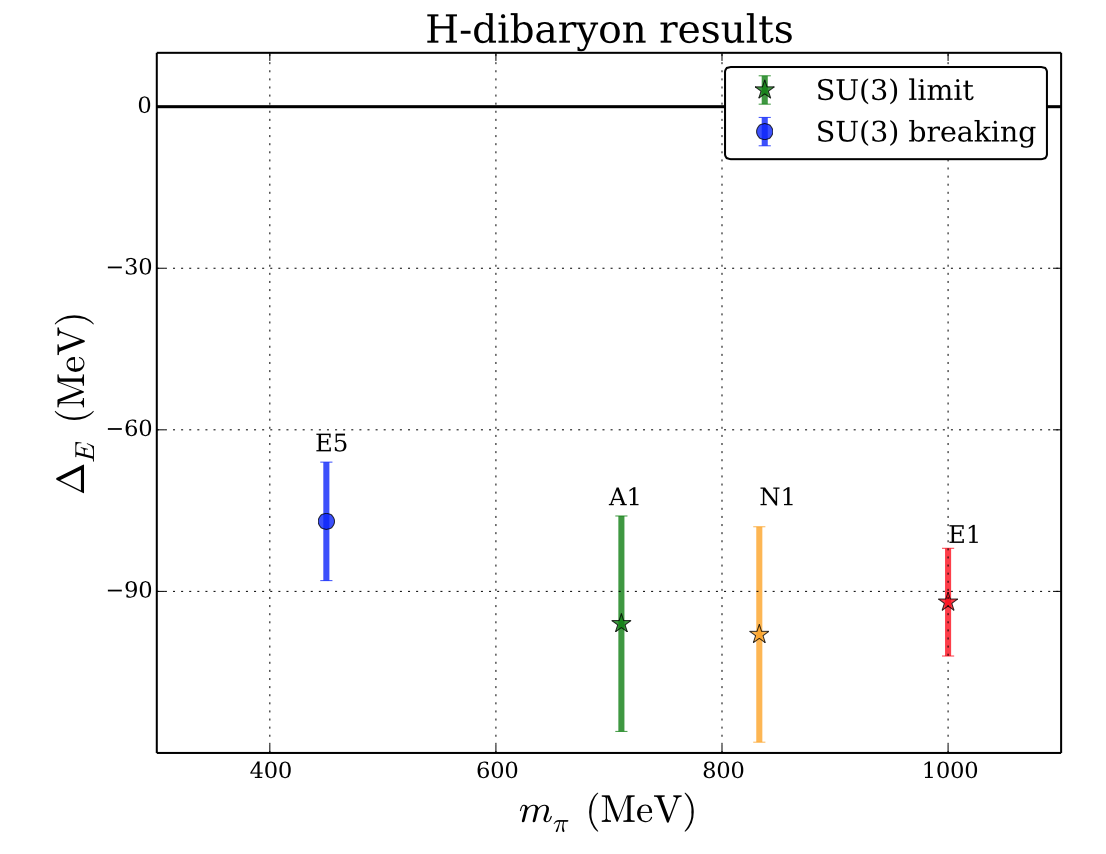}  
 \end{center}
\caption{
The binding energy of the H-dibaryon over a range of pion masses calculated by the Mainz Lattice Collaboration~\cite{Green:2014dea,Junnarkar:2015jyf}. 
\label{fig:Mainzsummary}
}
\end{figure}
Extrapolating the results of these calculations with an ERE to locate the bound states, 
they find that the H-dibaryon is bound  for $m_\pi \gsim 450~{\rm MeV}$, as shown in Figure~\ref{fig:Mainzsummary}.
They find binding energies that are somewhat deeper than those of other 
calculations~\cite{Beane:2010hg,Inoue:2010es}.

\begin{figure}
\begin{center}
 \includegraphics[width=0.75\linewidth]{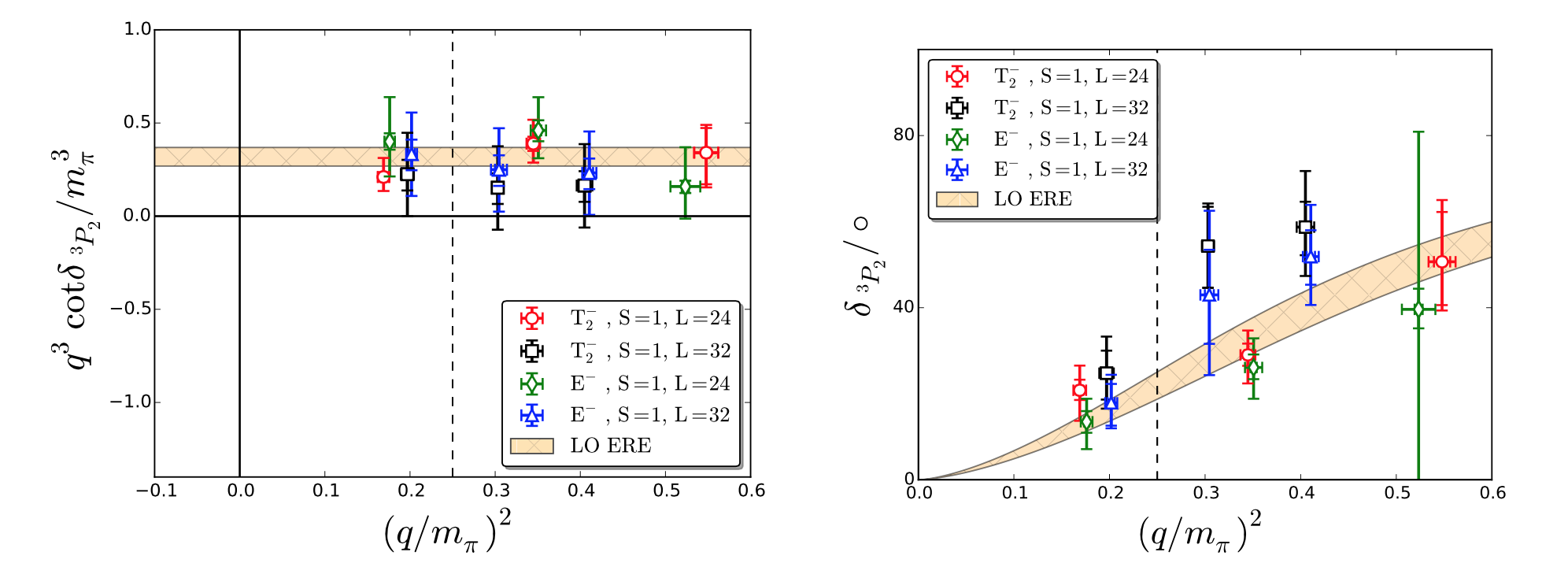}  
 \end{center}
\caption{
The left panel shows 
the results of CalLatt's calculation of
the real part of the inverse scattering amplitude in the $^3$P$_2$ channel, 
while the right panel shows the associated phase shift~\protect\cite{Berkowitz:2015eaa}.
\label{fig:pwaves}
}
\end{figure}

The CalLatt team has calculated p-wave and higher partial wave phase shifts~\cite{Berkowitz:2015eaa} 
on two ensembles of  NPLQCD gauge-field configurations with $m_\pi\sim 805~{\rm MeV}$, as shown in Figure~\ref{fig:pwaves}.
For the p-waves, they find non-zero phase shifts in the different cubic irreps. which agree within uncertainties,
and which are seen to be dominated by the scattering volume.
In the s-wave channels, they find binding energies that agree with the previous NPLQCD 
results~\cite{Beane:2012vq,Beane:2013br} within their uncertainties.  
In addition, using a truncated ERE, they find a state shifted by $\sim 3~{\rm MeV}$ in the deuteron channel, 
that is consistent with zero within uncertainties.  
It is likely that this state is an artifact of working with only two lattice volumes, 
unlike the three used by NPLQCD, and of their analysis methods.
Despite the defects associated with the suggestion of there being a second bound state in the deuteron 
channel at this pion mass, some have speculated that this is a cause for concern in using 
L\"usher's method to extract two-baryon 
information from LQCD calculations~\cite{Aoki:2016dmo}.  I believe this not to be the case.


%
\begin{figure}
\begin{center}
 \includegraphics[width=0.55\linewidth]{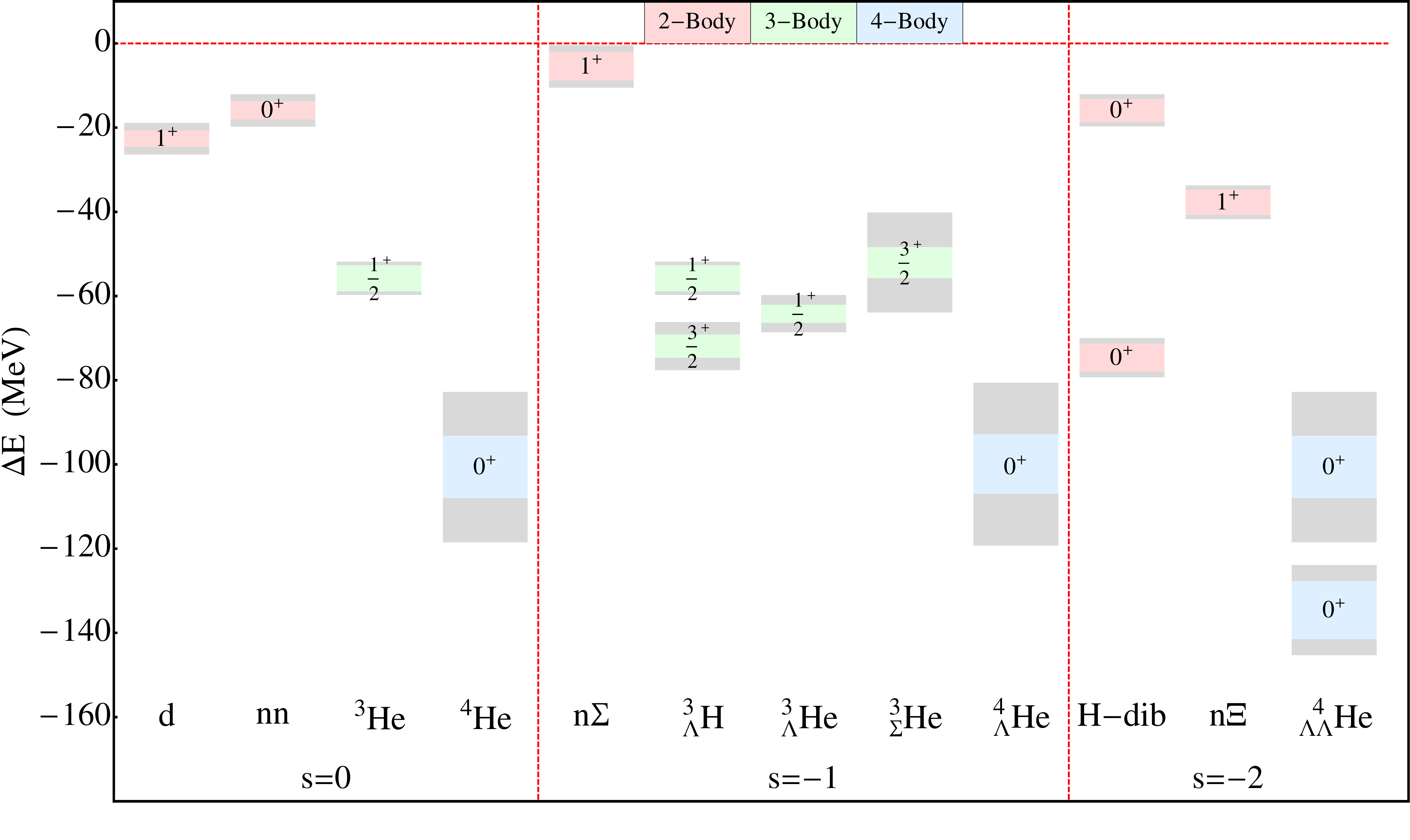}  
 \end{center}
\caption{
Low-lying states in s-shell nuclei and hypernuclei at a pion mass of 
$m_\pi\sim 805~{\rm MeV}$~\protect\cite{Beane:2012vq,Beane:2013br}.
\label{fig:NPLQCDnuclei}
}
\end{figure}
Since 2004, the NPLQCD collaboration has been calculating the properties and interactions of $A=2,3,4,5$ systems.
The first comprehensive analysis of light nuclei and hypernuclei at $m_\pi\sim 805~{\rm MeV}$ was performed in 
2011-2013~\cite{Beane:2012vq,Beane:2013br},
the results of which are shown in Figure~\ref{fig:NPLQCDnuclei}.
Generically, one finds that the nuclei are more deeply bound at heavier pion masses.
Surprisingly, the scattering parameters extracted from the phase-shift analysis, indicates that the deuteron 
remains a ``fluffy'' nucleus over a large range of pion masses and is unlikely to be fine-tuned.
It appears that this is a generic feature of a Yang-Mills theory with three ``light'' quarks.
The s-wave scattering phase shifts have been determined at $m_\pi\sim 805~{\rm MeV}$~\cite{Beane:2013br} and 
$m_\pi\sim 450~{\rm MeV}$~\cite{Orginos:2015aya} 
using L\"usher's method, and it is observed that the phase shifts exhibit zero-crossings, 
and  it is close to that of nature in both spin channels  at  $m_\pi\sim 450~{\rm MeV}$.  
The appropriate low-energy effective field theory (EFT) 
counterterms were constrained, permitting chiral extrapolations.
Hyperon-nucleon scattering has been investigated using both L\"usher's method and also by fitting the coefficients of a low-energy effective Hamiltonian at leading order in Weinberg's power counting~\cite{Beane:2006gf,Beane:2012ey} to the finite-volume energy eigenvalues.  
The later was done by explicitly 
diagonalizing the finite-volume Hamiltonian matrix to fit the energy eigenvalues, and then using this 
Hamiltonian to predict the continuum bound state energies and scattering amplitude~\cite{Beane:2012ey}.
The uncertainties associated with the extrapolated quantities are somewhat larger than experiment, and these are in the process of being reduced through further calculations.

\begin{figure}
\begin{center}
 \includegraphics[width=0.32\linewidth]{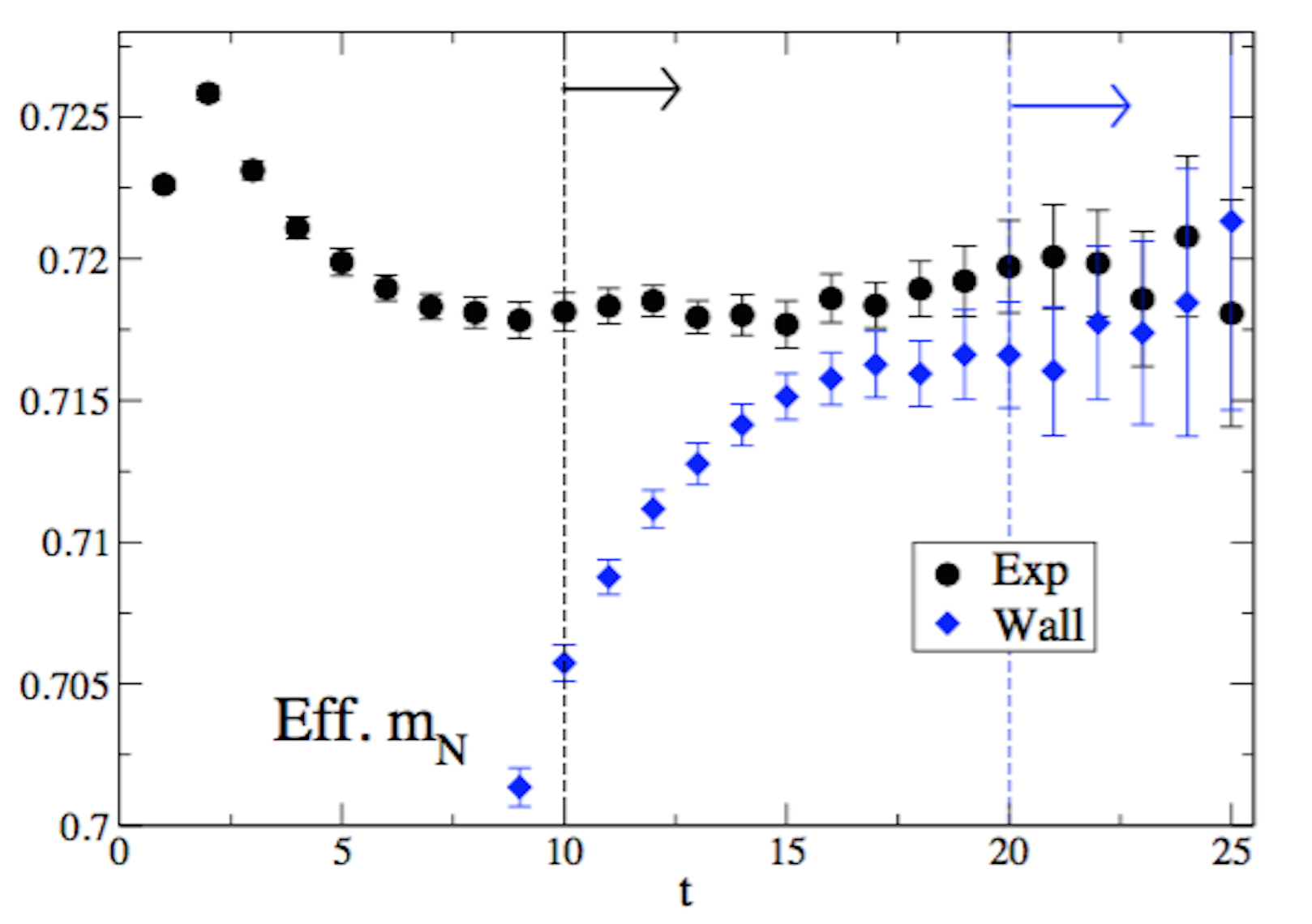}  
 \includegraphics[width=0.32\linewidth]{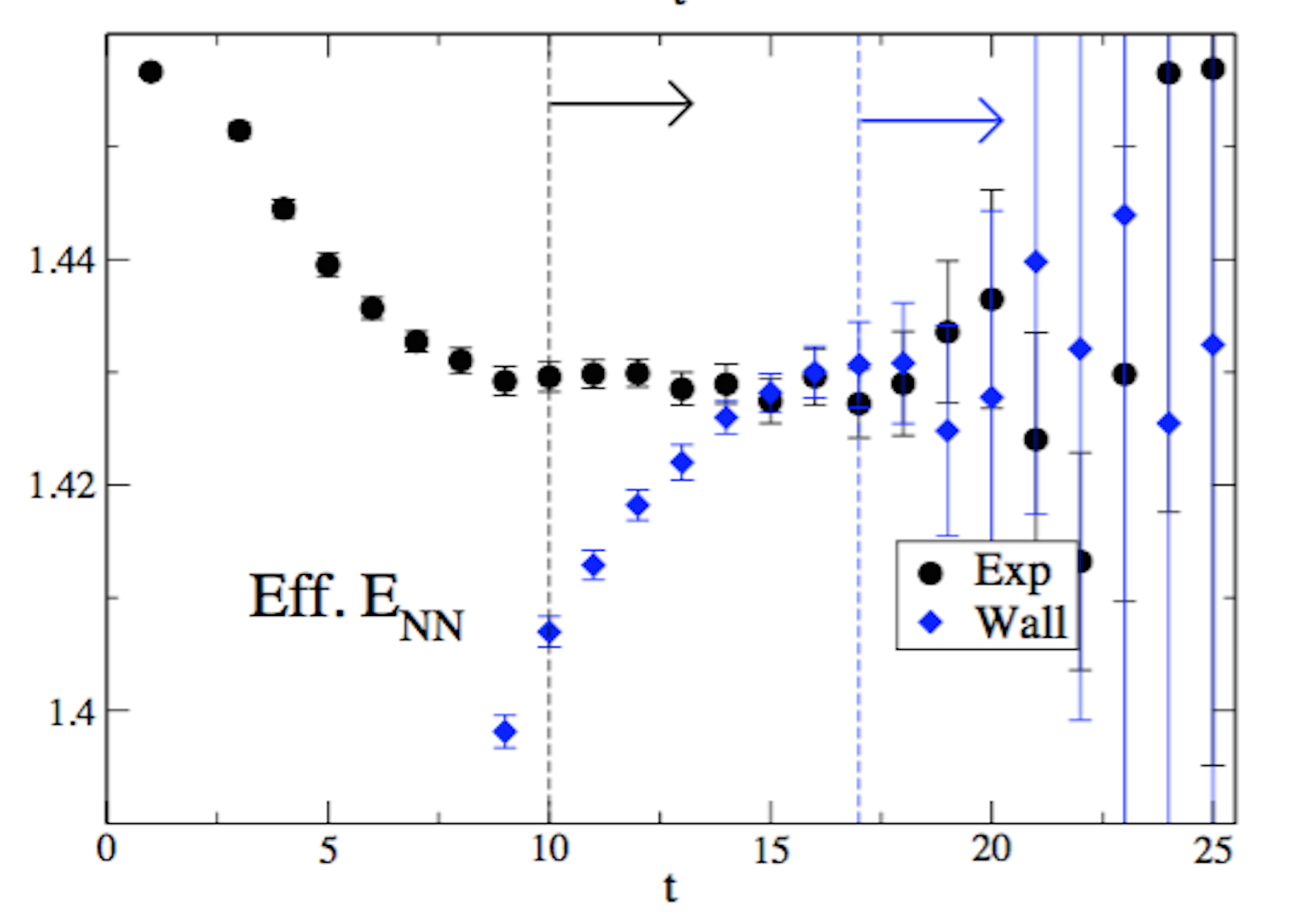}  
 \includegraphics[width=0.32\linewidth]{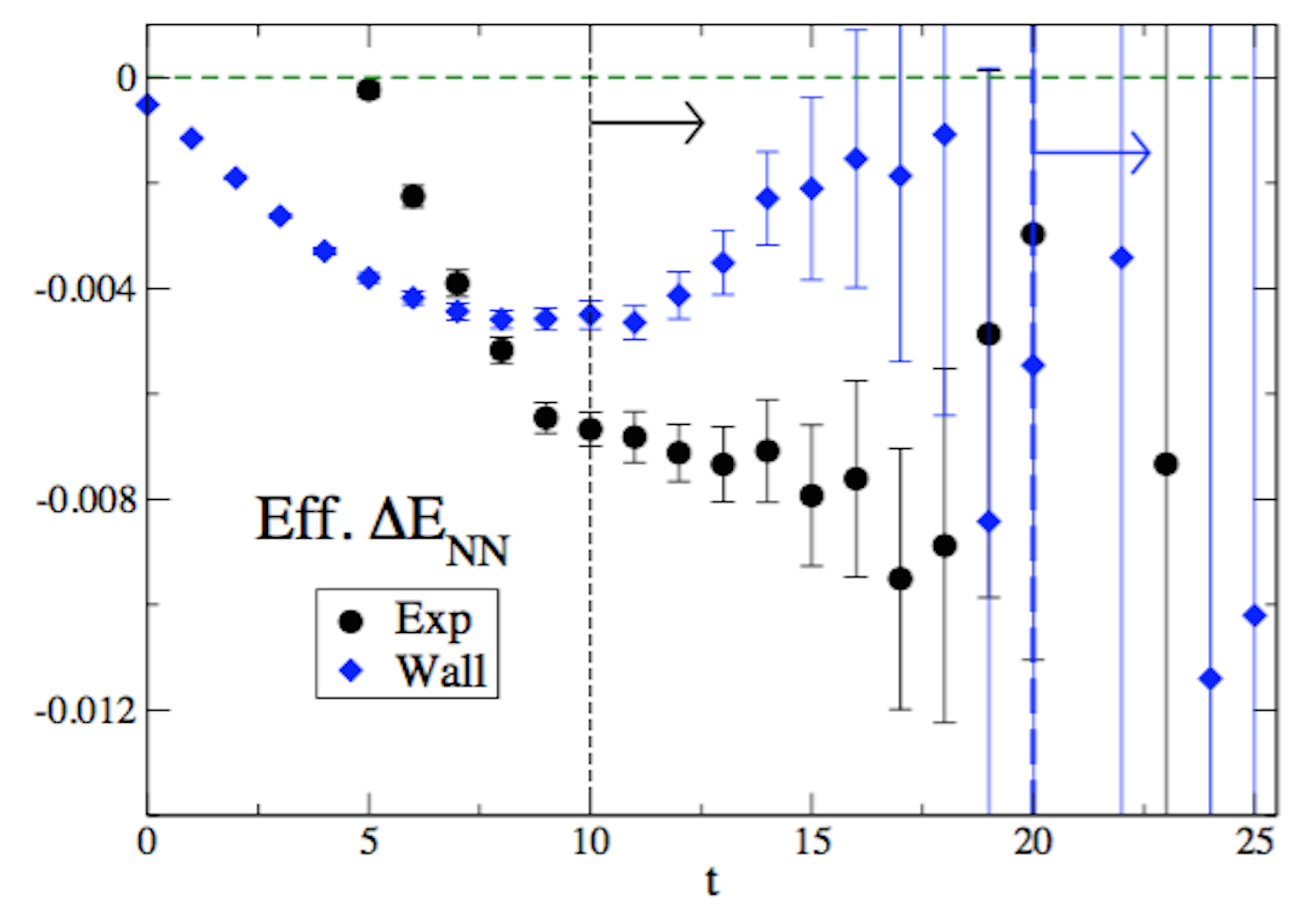}  
 \end{center}
\caption{
A comparison between correlation functions generated from wall-sources and exponentially-smeared sources by the PACS collaboration.
The left panel shows nucleon effective masses, the middle panel shows the two-nucleon effective energies and the right panel shows the 
binding energies formed from the ratio of correlation functions.
[I thank Takeshi Yamasaki for allowing me to show these clarifying figures.]
\label{fig:wallexpyama}
}
\end{figure}

There have been some recent suggestions of a ``Plateau Crisis'' by the HALQCD 
collaboration~\cite{Aoki:2016dmo,Iritani:2016jie}, 
suggesting that 
results obtained using 
L\"usher's method have identified false plateaus in their energy spectra.  
Let me make a few comments about such statements.
The implicit suggestion is that
the HALQCD method is superior and reliable because it does not require plateaus. 
My own experience suggests that the only ``crisis'' that has been encountered is not requiring plateaus in 
effective masses prior to extracting observables using the HALQCD method.
PACS  has reproduced the effect from correlation functions that are claimed to demonstrate the ``crisis'', 
as shown in Figure~\ref{fig:wallexpyama}.
As already discussed, the effective masses from the wall sources plateau much latter, as shown by the blue points.  
The binding energy formed from the ratios of wall-source correlation functions exhibits a false plateau in a time interval 
(around $t\sim 10$)
due to the significant contamination from  excited states, and  should be discarded.
Plateau'ing in both wall-source correlation functions occurs for 
$t\gsim 18$, and significantly more statistics are required to extract a statistically meaningful binding energy from those higher time-slices.
In contrast, the localized sources, such as the exponentially-smeared sources used to generate the black points in 
Figure~\ref{fig:wallexpyama}, provide plateaus in the effective masses at much earlier times.
A  plateau can be extracted from the ratio of correlation functions at an earlier time than from the wall-sources, 
from $t\gsim 12$, and is statistically significant.
Both  NPLQCD and PACS  employ smeared local sources for the quark propagators, 
and both the single-hadron and two-baryon energies  have plateaued in the time-slices 
from which the binding and continuum energies are determined. 
It would be helpful if collaborations showed  effective mass plots associated 
with the single-hadron and two-hadron 
systems, along with energy differences, in future publications.

\begin{figure}
\begin{center}
 \includegraphics[width=0.5\linewidth]{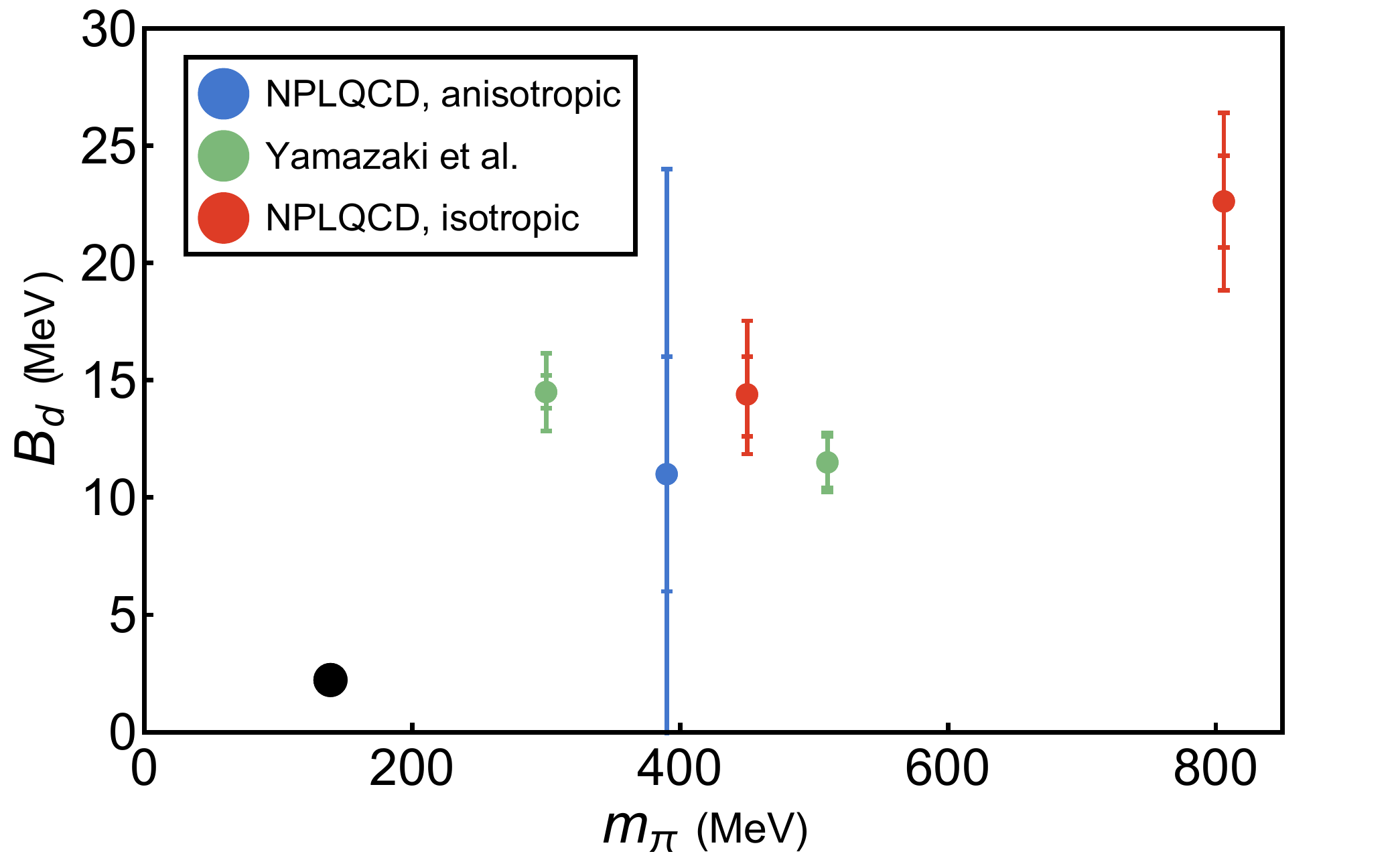}  
 \end{center}
\caption{
The deuteron binding energy as a function pion mass~\cite{Beane:2011iw,Beane:2012vq,Yamazaki:2015asa}.
In contract, HALQCD does not find a bound deuteron at the heavier pion masses.
\label{fig:deut}
}
\end{figure}
Due to its unnaturally small binding energy,
resulting from a delicate cancellation between short-, medium- and long-range physics,
calculating the deuteron binding energy at the physical point will be challenging, requiring large volumes and fine lattice spacings.
A compilation of independent calculations of the deuteron binding energy is presented in Figure~\ref{fig:deut}.
The number of calculations is small, and calculations over a range of pion masses is required to extrapolate to the physical point.

One of the exciting recent developments in the field is the first serious efforts to 
match the results of LQCD calculations to a low-energy EFT, and then predict the 
properties of nuclei beyond those of the lattice results~\cite{Barnea:2013uqa,Kirscher:2015tka}. 
\begin{figure}
\begin{center}
 \includegraphics[width=0.6\linewidth]{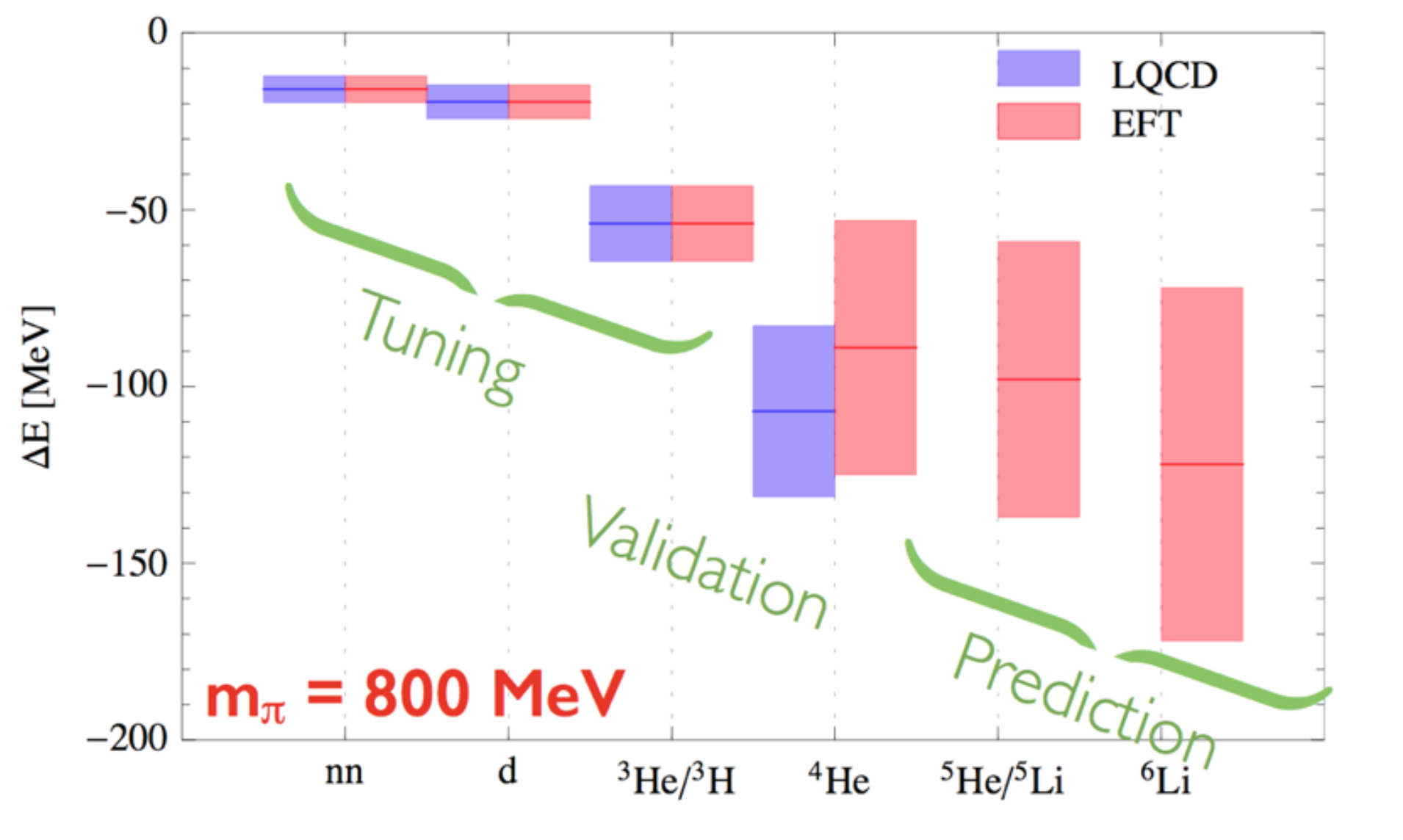}  
 \end{center}
\vskip -0.2in
\caption{
The two-nucleon and three-nucleon scattering parameters and binding energies calculated at $m_\pi\sim 805~{\rm MeV}$ 
were used to constrain interactions in the pionless EFT.  
Predictions for the $A=5,6$ systems are 
a prediction at this pion mass~\cite{Barnea:2013uqa,Kirscher:2015tka}.
\label{fig:EFTmatch}
}
\end{figure}
This is putting in place, albeit at unphysical quark masses, one of the critical components of the program to be able to make QCD predictions for elements of the Periodic Table.
They used the results of the two-nucleon and three-nucleon energy eigenvalues (scattering parameters and binding energies)
to constrain the two-nucleon and three-nucleon effective interactions in the pionless EFT, 
which is appropriate to use in the case of $m_\pi\sim 805~{\rm MeV}$ pions.
A comparison between the predicted four-nucleon binding energy and that calculated with the EFT showed that the four-nucleon interactions are small (within the uncertainty of the calculation) and verified the two nucleon and three-nucleon forces.
The EFT was then used to predict the binding of $^5$He, $^5$Li and $^6$Li at this pion mass, 
as shown in Figure~\ref{fig:EFTmatch}.
This is a major development in the field, and guides the way for connecting future LQCD calculations of multi-nucleon systems to elements far into the Periodic Table.

Another major recent development was the first LQCD calculation of an inelastic nuclear reaction cross section~\cite{Beane:2015yha}.
Using background magnetic fields, the NPLQCD collaboration calculated the low-energy cross section for the 
radiative capture process  $np\rightarrow d\gamma$, which  is dominated by the M1 amplitude in this energy regime.
\begin{figure}[!ht]
\begin{center}
 \includegraphics[width=0.43\linewidth]{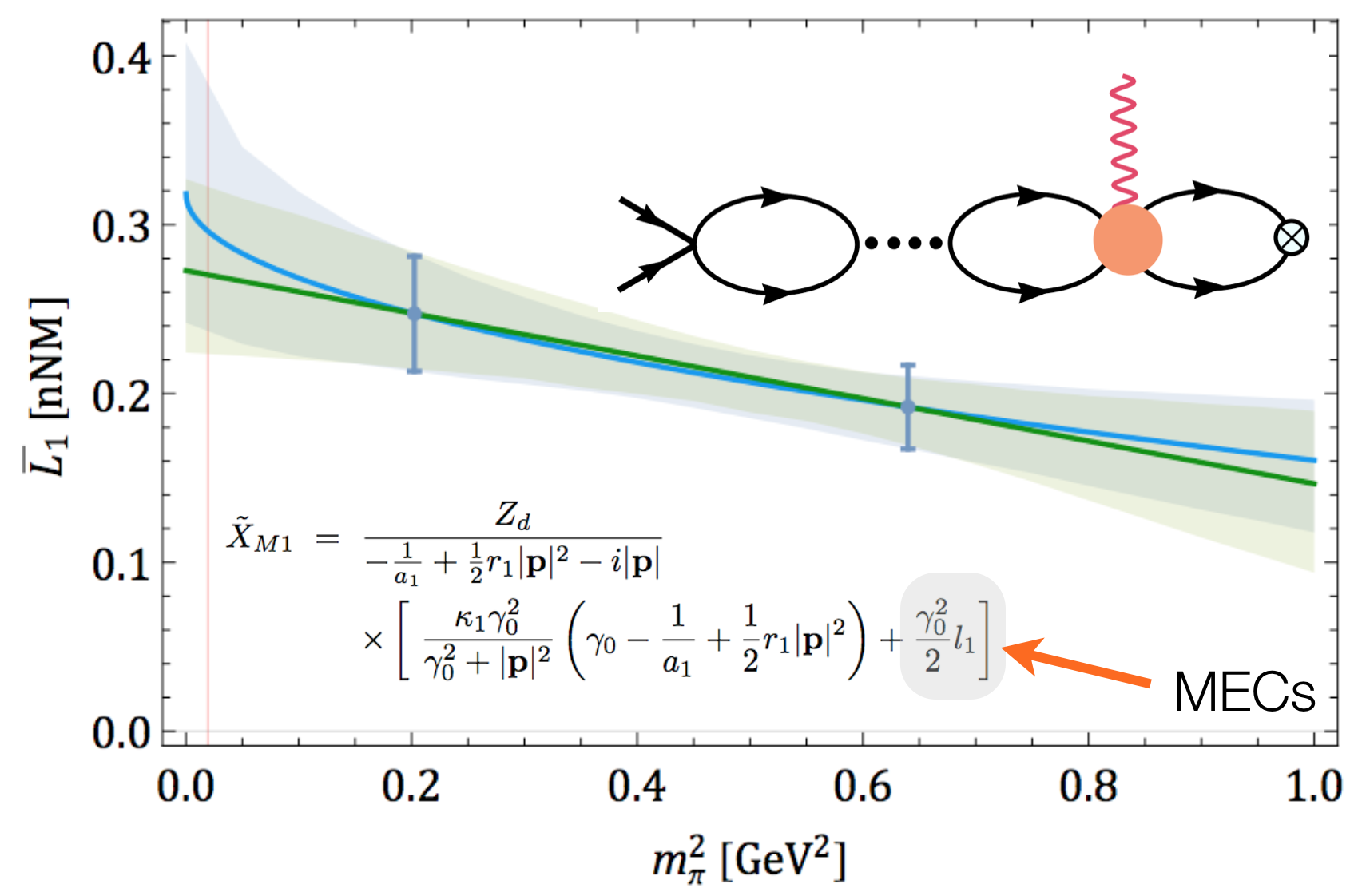}  
 \includegraphics[width=0.51\linewidth]{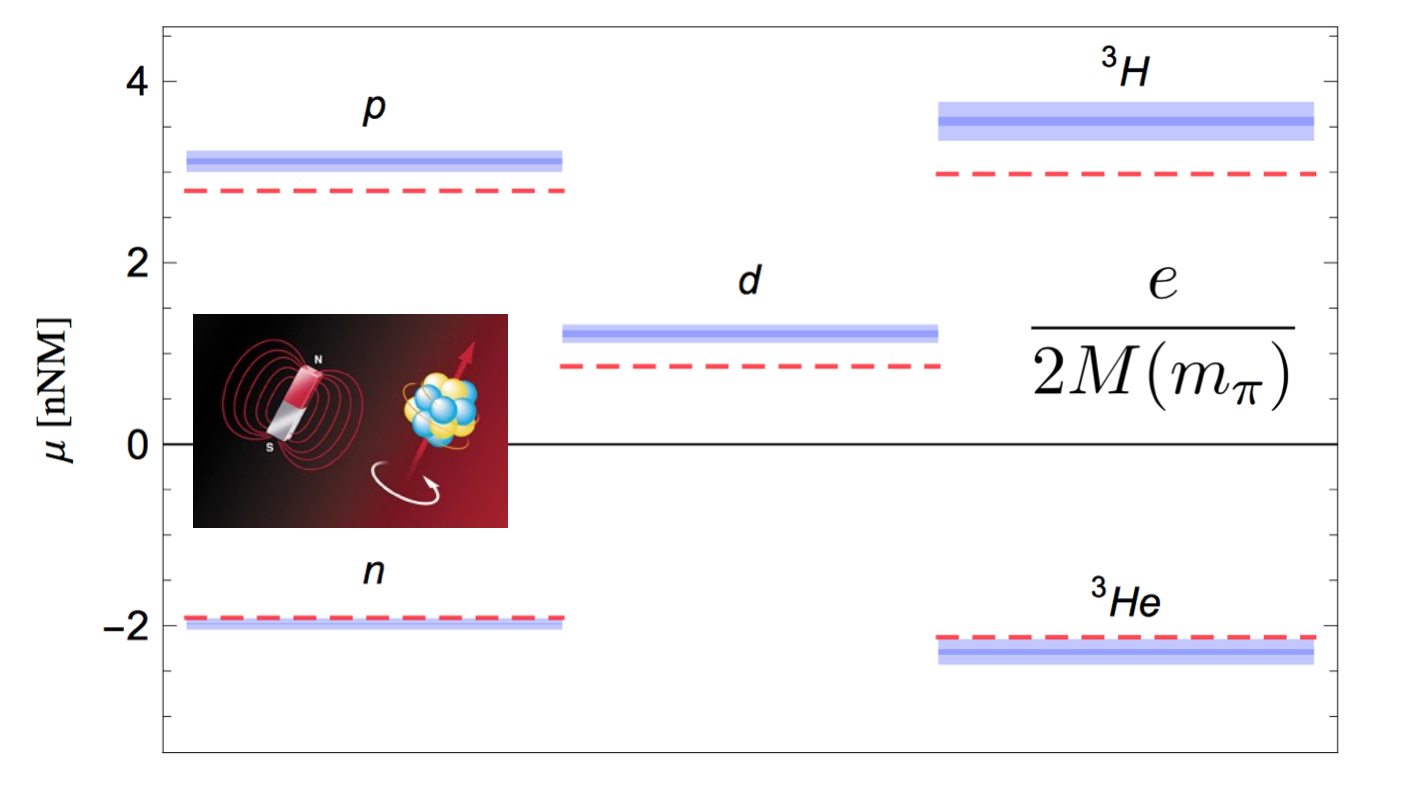}  
 \end{center}
\caption{
The correlated two-nucleon interaction (meson-exchange currents) extrapolated to the physical pion mass~\protect\cite{Beane:2015yha} (left panel), and the 
magnetic moments of the s-shell nuclei at $m_\pi\sim 805~{\rm MeV}$~\protect\cite{Beane:2014ora,Chang:2015qxa}.  
The red-dashed horizontal lines correspond to  experimental values.
\label{fig:L1nplqcd}
}
\end{figure}
Performing calculations 
of the energy splittings of two-nucleon systems in background magnetic fields 
at pion masses of $m_\pi\sim 805~{\rm MeV}$ and $m_\pi\sim 450~{\rm MeV}$, 
the NPLQCD collaboration isolated the correlated two-nucleon interaction in the pionless EFT, 
attributed to meson-exchange currents, and found only mild quark-mass dependence, similar to that 
observed for the magnetic moments, as shown in Figure~\ref{fig:L1nplqcd}.  
Using this quantity extrapolated to the physical pion mass, and the experimentally 
determine scattering parameters, a cross section of $\sigma^{\rm lccd}=334.9(5.3)~{\rm mb}$  was predicted at an incident neutron speed of $v=2,200~{\rm m/s}$, which is to be compared with the experimental cross section of $\sigma^{\rm lccd}=334.2(0.5)~{\rm mb}$.
The magnetic moments of the light nuclei have also been calculated and 
$m_\pi\sim 805~{\rm MeV}$~\protect\cite{Beane:2014ora,Chang:2015qxa}.
When expressed in units of natural Nuclear Magnetons, they agree remarkably well with the corresponding experimental values, 
as shown in Figure~\ref{fig:L1nplqcd}.


On a more exotic topic, it is conceivable that the dark matter in our universe are 
composite particles from a confining gauge theory.
It is then plausible that the dark matter is not simply single ``hadrons'' of this 
exotic gauge interaction, but also the nuclei that are likely to be created also.   
Interesting work in this area was done by Detmold and collaborators at MIT~\cite{Detmold:2014qqa,Detmold:2014kba}, in which they 
calculated the spectrum of light nuclei resulting from a SU(2) gauge group as a candidate for the dark matter.  They found some interesting results with regard to multi-scale dark matter with the possibility of inelastic reactions.


There are several calculations being pursued in multi-baryon systems that are of importance.
In the two-baryon systems, calculations of the binding energies are continuing at the physical point by PACS and HALQCD using distinct methods.  PACS is focused on nucleon-nucleon interactions from localized smeared sources  by direct calculation of the ground state energies, while HALQCD is deriving energy-dependent non-local interactions from wall-sources without requiring plateaus in 
effective mass plots.  
NPLQCD and the Mainz LQCD collaboration are pursuing calculations in multiple volumes at unphysical pion masses.  As the strange baryons have a less severe signal-to-noise problem than the nucleons, results in the  high-strangeness systems are more precise.
The progress in calculating electroweak matrix elements in multi-nucleon systems will continue, with first calculations of axial-current matrix elements expected in the near future, including those dictating the cross section for proton-proton fusion and the Gamow-Teller matrix element for tritium 
$\beta$-decay~\cite{Savage:2016kon}.  
One also expects to see progress in calculating matrix elements related to $\beta\beta$-decay of 
nuclei~\cite{Nicholson:2016byl}.

\section{Conclusions}

The field of nuclear physics is about to be revolutionized by the ability to reliably calculate low-energy strong interaction quantities using LQCD.  The low-lying spectra and simple properties of light nuclei, along with simple nuclear reactions, are now being calculated over a range of pion masses directly from QCD - which was unthinkable just 15 years ago.   With exascale computing resources arriving in the near future, precise calculations of light nuclei and their interactions from QCD, 
at the physical quark masses and including electromagnetism, will become straightforward,
and are critical to the success of the experimental programs in both high-energy physics and nuclear physics.
One of the really exciting developments witnessed during the last year or so is groups  working at the physical point, and one hopes that these efforts can accumulate adequate statistics to 
reproduce the experimental values of the  two-nucleon scattering parameters and the 
deuteron binding energy.

\acknowledgments

I would like to thank Takeshi Yamazaki, Takumi Doi and and Takumi Iritani for sending material that I included in my presentation and in these proceedings.
I am indebted to my collaborators in NPLQCD (Silas Beane, Emmanuel Chang, Zohreh Davoudi,Will Detmold, Kostas Orginos, Assumpta Parre\~no, Phiala Shanahan, Brian Tiburzi, Michael Wagman and Frank Winter).
I am supported  by DOE grant number~DE-FG02-00ER41132 and DE-SC00-10337, and  in part by the 
USQCD SciDAC project.
This research was supported in part by the National Science Foundation under grant number NSF PHY11-25915 and we thank the Kavli Institute for Theoretical Physics for hospitality 
during completion of these proceedings.

\bibliographystyle{JHEP} 
\bibliography{bib_Lattice2016.bib} 

\providecommand{\href}[2]{#2}\begingroup\raggedright\begin{thebibliography}{10}

\bibitem{Gandolfi:2015jma}
S.~Gandolfi, A.~Gezerlis and J.~Carlson, \emph{{Neutron Matter from Low to High
  Density}},
  \href{http://dx.doi.org/10.1146/annurev-nucl-102014-021957}{\emph{Ann. Rev.
  Nucl. Part. Sci.} {\bf 65} (2015) 303--328},
  [\href{http://arxiv.org/abs/1501.05675}{{\tt 1501.05675}}].

\bibitem{Hebeler:2013nza}
K.~Hebeler, J.~M. Lattimer, C.~J. Pethick and A.~Schwenk, \emph{{Equation of
  state and neutron star properties constrained by nuclear physics and
  observation}},
  \href{http://dx.doi.org/10.1088/0004-637X/773/1/11}{\emph{Astrophys. J.} {\bf
  773} (2013) 11}, [\href{http://arxiv.org/abs/1303.4662}{{\tt 1303.4662}}].

\bibitem{Abbott:2016blz}
{\scshape Virgo, LIGO Scientific} collaboration, B.~P. Abbott et~al.,
  \emph{{Observation of Gravitational Waves from a Binary Black Hole Merger}},
  \href{http://dx.doi.org/10.1103/PhysRevLett.116.061102}{\emph{Phys. Rev.
  Lett.} {\bf 116} (2016) 061102}, [\href{http://arxiv.org/abs/1602.03837}{{\tt
  1602.03837}}].

\bibitem{Mantysaari:2016ykx}
H.~Mäntysaari and B.~Schenke, \emph{{Evidence of strong proton shape
  fluctuations from incoherent diffraction}},
  \href{http://dx.doi.org/10.1103/PhysRevLett.117.052301}{\emph{Phys. Rev.
  Lett.} {\bf 117} (2016) 052301}, [\href{http://arxiv.org/abs/1603.04349}{{\tt
  1603.04349}}].

\bibitem{nsacLRP2015}
{Donald Geesaman}, \emph{{The 2015 Long Range Plan for Nuclear Science:
  Reaching for the Horizon}},  09, 2015.

\bibitem{Beane:2009kya}
S.~R. Beane, W.~Detmold, T.~C. Luu, K.~Orginos, A.~Parreno, M.~J. Savage
  et~al., \emph{{High Statistics Analysis using Anisotropic Clover Lattices:
  (I) Single Hadron Correlation Functions}},
  \href{http://dx.doi.org/10.1103/PhysRevD.79.114502}{\emph{Phys. Rev.} {\bf
  D79} (2009) 114502}, [\href{http://arxiv.org/abs/0903.2990}{{\tt
  0903.2990}}].

\bibitem{Detmold:2012eu}
W.~Detmold and K.~Orginos, \emph{{Nuclear correlation functions in lattice
  QCD}}, \href{http://dx.doi.org/10.1103/PhysRevD.87.114512}{\emph{Phys.Rev.}
  {\bf D87} (2013) 114512}, [\href{http://arxiv.org/abs/1207.1452}{{\tt
  1207.1452}}].

\bibitem{Doi:2012xd}
T.~Doi and M.~G. Endres, \emph{{Unified contraction algorithm for multi-baryon
  correlators on the lattice}},
  \href{http://dx.doi.org/10.1016/j.cpc.2012.09.004}{\emph{Comput. Phys.
  Commun.} {\bf 184} (2013) 117}, [\href{http://arxiv.org/abs/1205.0585}{{\tt
  1205.0585}}].

\bibitem{Yamazaki:2012hi}
T.~Yamazaki, K.-i. Ishikawa, Y.~Kuramashi and A.~Ukawa, \emph{{Helium nuclei,
  deuteron and dineutron in 2+1 flavor lattice QCD}},
  \href{http://dx.doi.org/10.1103/PhysRevD.86.074514}{\emph{Phys. Rev.} {\bf
  D86} (2012) 074514}, [\href{http://arxiv.org/abs/1207.4277}{{\tt
  1207.4277}}].

\bibitem{Luscher:1986pf}
M.~Luscher, \emph{{Volume Dependence of the Energy Spectrum in Massive Quantum
  Field Theories. 2. Scattering States}},
  \href{http://dx.doi.org/10.1007/BF01211097}{\emph{Commun.Math.Phys.} {\bf
  105} (1986) 153--188}.

\bibitem{Luscher:1990ux}
M.~Luscher, \emph{{Two particle states on a torus and their relation to the
  scattering matrix}},
  \href{http://dx.doi.org/10.1016/0550-3213(91)90366-6}{\emph{Nucl.Phys.} {\bf
  B354} (1991) 531--578}.

\bibitem{HALQCD:2012aa}
{\scshape HAL QCD} collaboration, N.~Ishii, S.~Aoki, T.~Doi, T.~Hatsuda,
  Y.~Ikeda, T.~Inoue et~al., \emph{{Hadron?hadron interactions from
  imaginary-time Nambu?Bethe?Salpeter wave function on the lattice}},
  \href{http://dx.doi.org/10.1016/j.physletb.2012.04.076}{\emph{Phys. Lett.}
  {\bf B712} (2012) 437--441}, [\href{http://arxiv.org/abs/1203.3642}{{\tt
  1203.3642}}].

\bibitem{Beane:2013br}
{\scshape NPLQCD Collaboration} collaboration, S.~Beane et~al.,
  \emph{{Nucleon-Nucleon Scattering Parameters in the Limit of SU(3) Flavor
  Symmetry}},
  \href{http://dx.doi.org/10.1103/PhysRevC.88.024003}{\emph{Phys.Rev.} {\bf
  C88} (2013) 024003}, [\href{http://arxiv.org/abs/1301.5790}{{\tt
  1301.5790}}].

\bibitem{Davoudi:2011md}
Z.~Davoudi and M.~J. Savage, \emph{{Improving the Volume Dependence of Two-Body
  Binding Energies Calculated with Lattice QCD}},
  \href{http://dx.doi.org/10.1103/PhysRevD.84.114502}{\emph{Phys. Rev.} {\bf
  D84} (2011) 114502}, [\href{http://arxiv.org/abs/1108.5371}{{\tt
  1108.5371}}].

\bibitem{Briceno:2013lba}
R.~A. Briceno, Z.~Davoudi and T.~C. Luu, \emph{{Two-Nucleon Systems in a Finite
  Volume: (I) Quantization Conditions}},
  \href{http://dx.doi.org/10.1103/PhysRevD.88.034502}{\emph{Phys. Rev.} {\bf
  D88} (2013) 034502}, [\href{http://arxiv.org/abs/1305.4903}{{\tt
  1305.4903}}].

\bibitem{Briceno:2013hya}
R.~A. Briceno, Z.~Davoudi, T.~C. Luu and M.~J. Savage, \emph{{Two-Baryon
  Systems with Twisted Boundary Conditions}},
  \href{http://dx.doi.org/10.1103/PhysRevD.89.074509}{\emph{Phys.Rev.} {\bf
  D89} (2014) 074509}, [\href{http://arxiv.org/abs/1311.7686}{{\tt
  1311.7686}}].

\bibitem{Briceno:2013bda}
R.~A. Briceno, Z.~Davoudi, T.~Luu and M.~J. Savage, \emph{{Two-nucleon systems
  in a finite volume. II. $^3S_1-^3D_1$ coupled channels and the deuteron}},
  \href{http://dx.doi.org/10.1103/PhysRevD.88.114507}{\emph{Phys.Rev.} {\bf
  D88} (2013) 114507}, [\href{http://arxiv.org/abs/1309.3556}{{\tt
  1309.3556}}].

\bibitem{Yamazaki:2015asa}
T.~Yamazaki, K.-i. Ishikawa, Y.~Kuramashi and A.~Ukawa, \emph{{Study of quark
  mass dependence of binding energy for light nuclei in 2+1 flavor lattice
  QCD}}, \href{http://dx.doi.org/10.1103/PhysRevD.92.014501}{\emph{Phys. Rev.}
  {\bf D92} (2015) 014501}, [\href{http://arxiv.org/abs/1502.04182}{{\tt
  1502.04182}}].

\bibitem{Beane:2003da}
S.~R. Beane, P.~F. Bedaque, A.~Parreno and M.~J. Savage, \emph{{Two nucleons on
  a lattice}},
  \href{http://dx.doi.org/10.1016/j.physletb.2004.02.007}{\emph{Phys. Lett.}
  {\bf B585} (2004) 106--114},
  [\href{http://arxiv.org/abs/hep-lat/0312004}{{\tt hep-lat/0312004}}].

\bibitem{Beane:2012vq}
S.~Beane, E.~Chang, S.~Cohen, W.~Detmold, H.~Lin et~al., \emph{{Light Nuclei
  and Hypernuclei from Quantum Chromodynamics in the Limit of SU(3) Flavor
  Symmetry}},
  \href{http://dx.doi.org/10.1103/PhysRevD.87.034506}{\emph{Phys.Rev.} {\bf
  D87} (2013) 034506}, [\href{http://arxiv.org/abs/1206.5219}{{\tt
  1206.5219}}].

\bibitem{Orginos:2015aya}
K.~Orginos, A.~Parreno, M.~J. Savage, S.~R. Beane, E.~Chang and W.~Detmold,
  \emph{{Two nucleon systems at $m_\pi\sim 450~{\rm MeV}$ from lattice QCD}},
  \href{http://dx.doi.org/10.1103/PhysRevD.92.114512}{\emph{Phys. Rev.} {\bf
  D92} (2015) 114512}, [\href{http://arxiv.org/abs/1508.07583}{{\tt
  1508.07583}}].

\bibitem{Sasaki:2016gpc}
K.~Sasaki et~al., \emph{{First results of baryon interactions from lattice QCD
  with physical masses (3) -- Strangeness $S=-2$ two-baryon system}},
  {\emph{PoS} {\bf LATTICE2015} (2016) 088}.

\bibitem{Ishii:2016zsf}
N.~Ishii et~al., \emph{{First results of baryon interactions from lattice QCD
  with physical masses (2) -- S=-3 and S=-4 sectors ($\Xi\Xi, \Xi\Sigma,
  \Xi\Lambda-\Xi\Sigma$ channels) --}}, {\emph{PoS} {\bf LATTICE2015} (2016)
  087}.

\bibitem{Doi:2011gq}
{\scshape HAL QCD} collaboration, T.~Doi, S.~Aoki, T.~Hatsuda, Y.~Ikeda,
  T.~Inoue, N.~Ishii et~al., \emph{{Exploring Three-Nucleon Forces in Lattice
  QCD}}, \href{http://dx.doi.org/10.1143/PTP.127.723}{\emph{Prog. Theor. Phys.}
  {\bf 127} (2012) 723--738}, [\href{http://arxiv.org/abs/1106.2276}{{\tt
  1106.2276}}].

\bibitem{Iritani:2016jie}
T.~Iritani et~al., \emph{{Mirage in Temporal Correlation functions for
  Baryon-Baryon Interactions in Lattice QCD}},
  \href{http://dx.doi.org/10.1007/JHEP10(2016)101}{\emph{JHEP} {\bf 10} (2016)
  101}, [\href{http://arxiv.org/abs/1607.06371}{{\tt 1607.06371}}].

\bibitem{Green:2014dea}
J.~Green, A.~Francis, P.~Junnarkar, C.~Miao, T.~Rae and H.~Wittig,
  \emph{{Search for a bound H-dibaryon using local six-quark interpolating
  operators}}, {\emph{PoS} {\bf LATTICE2014} (2014) 107},
  [\href{http://arxiv.org/abs/1411.1643}{{\tt 1411.1643}}].

\bibitem{Junnarkar:2015jyf}
P.~Junnarkar, A.~Francis, J.~Green, C.~Miao, T.~Rae and H.~Wittig,
  \emph{{Search for the H-Dibaryon in two flavor Lattice QCD}}, {\emph{PoS}
  {\bf CD15} (2015) 079}, [\href{http://arxiv.org/abs/1511.01849}{{\tt
  1511.01849}}].

\bibitem{Beane:2010hg}
{\scshape NPLQCD} collaboration, S.~R. Beane et~al., \emph{{Evidence for a
  Bound H-dibaryon from Lattice QCD}},
  \href{http://dx.doi.org/10.1103/PhysRevLett.106.162001}{\emph{Phys. Rev.
  Lett.} {\bf 106} (2011) 162001}, [\href{http://arxiv.org/abs/1012.3812}{{\tt
  1012.3812}}].

\bibitem{Inoue:2010es}
{\scshape HAL QCD} collaboration, T.~Inoue, N.~Ishii, S.~Aoki, T.~Doi,
  T.~Hatsuda, Y.~Ikeda et~al., \emph{{Bound H-dibaryon in Flavor SU(3) Limit of
  Lattice QCD}},
  \href{http://dx.doi.org/10.1103/PhysRevLett.106.162002}{\emph{Phys. Rev.
  Lett.} {\bf 106} (2011) 162002}, [\href{http://arxiv.org/abs/1012.5928}{{\tt
  1012.5928}}].

\bibitem{Berkowitz:2015eaa}
E.~Berkowitz, T.~Kurth, A.~Nicholson, B.~Joo, E.~Rinaldi, M.~Strother et~al.,
  \emph{{Two-Nucleon Higher Partial-Wave Scattering from Lattice QCD}},
  \href{http://arxiv.org/abs/1508.00886}{{\tt 1508.00886}}.

\bibitem{Aoki:2016dmo}
S.~Aoki, T.~Doi and T.~Iritani, \emph{{L\"uscher's finite volume test for
  two-baryon systems with attractive interactions}},  2016.
\newblock \href{http://arxiv.org/abs/1610.09763}{{\tt 1610.09763}}.

\bibitem{Beane:2006gf}
{\scshape NPLQCD} collaboration, S.~R. Beane, P.~F. Bedaque, T.~C. Luu,
  K.~Orginos, E.~Pallante, A.~Parreno et~al., \emph{{Hyperon-Nucleon Scattering
  from Fully-Dynamical Lattice QCD}},
  \href{http://dx.doi.org/10.1016/j.nuclphysa.2007.07.006}{\emph{Nucl. Phys.}
  {\bf A794} (2007) 62--72}, [\href{http://arxiv.org/abs/hep-lat/0612026}{{\tt
  hep-lat/0612026}}].

\bibitem{Beane:2012ey}
S.~R. Beane, E.~Chang, S.~D. Cohen, W.~Detmold, H.~W. Lin, T.~C. Luu et~al.,
  \emph{{Hyperon-Nucleon Interactions and the Composition of Dense Nuclear
  Matter from Quantum Chromodynamics}},
  \href{http://dx.doi.org/10.1103/PhysRevLett.109.172001}{\emph{Phys. Rev.
  Lett.} {\bf 109} (2012) 172001}, [\href{http://arxiv.org/abs/1204.3606}{{\tt
  1204.3606}}].

\bibitem{Beane:2011iw}
{\scshape NPLQCD Collaboration} collaboration, S.~Beane et~al., \emph{{The
  Deuteron and Exotic Two-Body Bound States from Lattice QCD}},
  \href{http://dx.doi.org/10.1103/PhysRevD.85.054511}{\emph{Phys.Rev.} {\bf
  D85} (2012) 054511}, [\href{http://arxiv.org/abs/1109.2889}{{\tt
  1109.2889}}].

\bibitem{Barnea:2013uqa}
N.~Barnea, L.~Contessi, D.~Gazit, F.~Pederiva and U.~van Kolck,
  \emph{{Effective Field Theory for Lattice Nuclei}},
  \href{http://dx.doi.org/10.1103/PhysRevLett.114.052501}{\emph{Phys. Rev.
  Lett.} {\bf 114} (2015) 052501}, [\href{http://arxiv.org/abs/1311.4966}{{\tt
  1311.4966}}].

\bibitem{Kirscher:2015tka}
J.~Kirscher, \emph{{Matching effective few-nucleon theories to QCD}},
  \href{http://dx.doi.org/10.1142/S0218301316410019}{\emph{Int. J. Mod. Phys.}
  {\bf E25} (2016) 1641001}, [\href{http://arxiv.org/abs/1509.07697}{{\tt
  1509.07697}}].

\bibitem{Beane:2015yha}
{\scshape NPLQCD} collaboration, S.~R. Beane, E.~Chang, W.~Detmold, K.~Orginos,
  A.~Parreño, M.~J. Savage et~al., \emph{{Ab initio Calculation of the
  $np\rightarrow d\gamma$ Radiative Capture Process}},
  \href{http://dx.doi.org/10.1103/PhysRevLett.115.132001}{\emph{Phys. Rev.
  Lett.} {\bf 115} (2015) 132001}, [\href{http://arxiv.org/abs/1505.02422}{{\tt
  1505.02422}}].

\bibitem{Beane:2014ora}
S.~R. Beane, E.~Chang, S.~Cohen, W.~Detmold, H.~W. Lin, K.~Orginos et~al.,
  \emph{{Magnetic moments of light nuclei from lattice quantum
  chromodynamics}},
  \href{http://dx.doi.org/10.1103/PhysRevLett.113.252001}{\emph{Phys. Rev.
  Lett.} {\bf 113} (2014) 252001}, [\href{http://arxiv.org/abs/1409.3556}{{\tt
  1409.3556}}].

\bibitem{Chang:2015qxa}
{\scshape NPLQCD} collaboration, E.~Chang, W.~Detmold, K.~Orginos, A.~Parreno,
  M.~J. Savage, B.~C. Tiburzi et~al., \emph{{Magnetic structure of light nuclei
  from lattice QCD}},
  \href{http://dx.doi.org/10.1103/PhysRevD.92.114502}{\emph{Phys. Rev.} {\bf
  D92} (2015) 114502}, [\href{http://arxiv.org/abs/1506.05518}{{\tt
  1506.05518}}].

\bibitem{Detmold:2014qqa}
W.~Detmold, M.~McCullough and A.~Pochinsky, \emph{{Dark Nuclei I: Cosmology and
  Indirect Detection}},
  \href{http://dx.doi.org/10.1103/PhysRevD.90.115013}{\emph{Phys. Rev.} {\bf
  D90} (2014) 115013}, [\href{http://arxiv.org/abs/1406.2276}{{\tt
  1406.2276}}].

\bibitem{Detmold:2014kba}
W.~Detmold, M.~McCullough and A.~Pochinsky, \emph{{Dark nuclei. II. Nuclear
  spectroscopy in two-color QCD}},
  \href{http://dx.doi.org/10.1103/PhysRevD.90.114506}{\emph{Phys. Rev.} {\bf
  D90} (2014) 114506}, [\href{http://arxiv.org/abs/1406.4116}{{\tt
  1406.4116}}].

\bibitem{Savage:2016kon}
M.~J. Savage, P.~E. Shanahan, B.~C. Tiburzi, M.~L. Wagman, F.~Winter, S.~R.
  Beane et~al., \emph{{Proton-proton fusion and tritium $\beta$-decay from
  lattice quantum chromodynamics}},
  \href{http://arxiv.org/abs/1610.04545}{{\tt 1610.04545}}.

\bibitem{Nicholson:2016byl}
A.~Nicholson, E.~Berkowitz, C.~C. Chang, M.~A. Clark, B.~Joo, T.~Kurth et~al.,
  \emph{{Neutrinoless double beta decay from lattice QCD}},  in
  \emph{{Proceedings, 34th International Symposium on Lattice Field Theory
  (Lattice 2016): Southampton, UK, July 24-30, 2016}}, 2016.
\newblock \href{http://arxiv.org/abs/1608.04793}{{\tt 1608.04793}}.

\end{thebibliography}\endgroup

\end{document}